\documentclass[10pt,twocolumn,reprint,amsmath,amssymb,aps,pra,showpacs,longbibliography,superscriptaddress]{revtex4-1}
\usepackage[latin9]{inputenc}
\usepackage{amsmath}
\usepackage{amssymb}
\usepackage{graphicx}

\makeatletter

\newcommand*\LyXThinSpace{\,\hspace{0pt}}

\usepackage{amsfonts}
\usepackage{graphicx}
\usepackage{graphics}

\makeatother

\begin{document}
\title{Second sound with ultracold atoms: A brief historical account}
\author{Hui Hu}
\affiliation{Centre for Quantum Technology Theory, Swinburne University of Technology,
Melbourne 3122, Australia}
\author{Xing-Can Yao}
\affiliation{Hefei National Research Center for Physical Sciences at the Microscale
and School of Physical Sciences, University of Science and Technology
of China, Hefei 230026, China}
\affiliation{Shanghai Research Center for Quantum Science and CAS Center for Excellence
in Quantum Information and Quantum Physics, University of Science
and Technology of China, Shanghai 201315, China }
\affiliation{Hefei National Laboratory, University of Science and Technology of
China, Hefei 230088, China}
\author{Xia-Ji Liu}
\affiliation{Centre for Quantum Technology Theory, Swinburne University of Technology,
Melbourne 3122, Australia}
\affiliation{Kavli Institute for Theoretical Physics, UC Santa Barbara, USA}
\date{\today}
\begin{abstract}
We briefly review the research on second sound in ultracold atomic
physics, with emphasis on strongly interacting unitary Fermi gases
with infinitely large $s$-wave scattering length. Second sound is
a smoking-gun feature of superfluidity in any quantum superfluids.
The observation and characterization of second sound in ultracold
quantum gases has been a long-standing challenge, and in recent years
there are rapid developments due to the experimental realization of
a uniform box-trap potential. The purpose of this review is to present
a brief historical account of the key research activities on second
sound over the past two decades. We summarize the initial theoretical
works that reveal the characteristics of second sound in a unitary
Fermi gas, and introduce its first observation in a highly elongated
harmonic trap. We then discuss the most recent measurement on second
sound attenuation in a uniform setup, which may open a new era to
understand quantum transport near quantum criticality in the strongly
interacting regime. The observation of second sound in homogeneous
weakly interacting Bose condensates in both two and three dimensions
are also briefly introduced.
\end{abstract}
\maketitle

\section{Introduction}

Second sound is a temperature or entropy wave that propagates inside
a quantum superfluid, contrasted to the usual density wave of first
sound \citep{Donnelly2009PhysicsToday}. It is the hallmark of superfluidity:
the second sound velocity directly relates to the superfluid fraction,
which measures the number of particles undergoing quantum frictionless
motion without costing energy; while the second sound attenuation
links to the transport coefficients that characterize the momentum
and heat transport \citep{Donnelly2009PhysicsToday}. Due to the crucial
role of second sound played in quantum superfluids, its observation
and characterization is one of the most important research goals in
the studies of ultracold atomic quantum gases \citep{SecondSoundPSReviewarXiv}.

For a strongly interacting unitary Fermi gas with a divergent $s$-wave
scattering length ($a=\infty$ or $1/a=0$) \citep{Bloch2008RMP},
the search for second sound started soon after its realization in
2002 \citep{OHara2002Science}. By using a broad Feshbach resonance
located near $B\simeq832$ G \citep{Chin2010RMP}, K. M. O\textquoteright Hara
and his colleagues at Duke University successfully demonstrated the
ballistic hydrodynamic expansion of a Fermi cloud of lithium atoms,
due to either the superfluid hydrodynamics or normal collisional hydrodynamics
\citep{OHara2002Science}. This observation paves the way to second
sound, if the superfluid hydrodynamics is responsible for the ballistic
expansion. Collective density oscillations (i.e., first sounds) of
the strongly interacting Fermi gas in harmonic traps were subsequently
measured \citep{Kinast2004PRL,Bartenstein2004PRL,Altmeyer2007PRL}.
The resulting breathing mode frequency was found to be well explained
by the superfluid hydrodynamic theory \citep{Stringari2004EPL,Hu2004PRL,Astrakharchik2005PRL}.
However, there was no trace of second sound in the measurement of
collective density excitations. Theoretical support was not favorable
either, since one needs to solve Landau's two-fluid hydrodynamic equations
in the presence of a harmonic trapping potential, which turns out
to be fairly non-trivial \citep{Shenoy1998PRL}.

This theoretical technical problem was partly solved by Allan Griffin
and his co-workers \citep{Taylor2005PRA,Taylor2008PRA,Taylor2009PRA},
by reformulating the dissipationless two-fluid hydrodynamic theory
into a variational form \citep{Taylor2005PRA}. As inspired by the
successful confirmation of Fermi superfluidity via the measurements
of condensate fraction \citep{Regal2004PRL,Zwierlein2004PRL} and
vortex lattices \citep{Zwierlein2005Nature}, more accurate predictions
were obtained for the superfluid density \citep{Fukushima2007PRA}
and equations of state \citep{Hu2006EPL}, within the framework of
the Gaussian pair fluctuation theory \citep{Hu2006EPL}. These predictions
provided a useful input to the variational solution of the two-fluid
hydrodynamic equations \citep{Taylor2008PRA}. As a result, various
hydrodynamic modes in an isotropic harmonic trap were solved and classified
as the first sound and second sound \citep{Taylor2009PRA}. A major
encouraging outcome of this analysis was that, the Landau-Placzek
ratio, which characterizes the coupling between first and second sounds
was found to be significant for a unitary Fermi gas. Therefore, in
principle one should be able to find the signal of second sound in
the density response, by fine tuning the density perturbation close
to the avoided crossing of first and second sounds \citep{Taylor2009PRA}.

Experimentally, however, it is difficult to construct a perfectly
spherical harmonic trap. For a general axially symmetric harmonic
trap, the variational solution of the two-fluid hydrodynamic equations
is not available, since too many variational parameters are needed
to obtain a convergent solution. To overcome this difficulty, Sandro
Stringari and his collaborators proposed to consider a highly-elongated
cigar trap, for which it is plausible to reduce the three-dimensional
two-fluid hydrodynamic equations into an effective one-dimensional
form \citep{Bertaina2010PRL,Hou2013PRA}. This brilliant idea turns
out to be very successful and eventually leads to the first observation
of the second sound propagation in a highly-elongated unitary Fermi
gas by L. A. Sidorenkov and his colleagues at the University of Innsbruck
in 2013 \citep{Sidorenkov2013Nature}. The temperature dependence
of superfluid density of the unitary Fermi gas was extracted for the
first time. However, the second sound attenuation and the related
transport coefficients remain undetermined. This is partly because
the reduced one-dimensional hydrodynamic equations are dissipationless
\citep{Bertaina2010PRL,Hou2013PRA}. It is not clear how to extract
the transport coefficients, even if one can measure the sound attenuation
in highly-elongated harmonic traps.

Therefore, harmonic trap potential is very unfriendly to the second
sound measurement. Remarkably, recent technical advances enable the
realization of a uniform box-trap potential \citep{Gaunt2013PRL}.
The experimental manipulation of a homogeneous unitary Fermi gas has
now been reported by several groups \citep{Mukherjee2017PRL,Hueck2018PRL}.
In addition to this technical advance, new knowledge of the transport
coefficients is also obtained. The shear viscosity of the unitary
Fermi gas has been both measured via ballistic expansion \citep{Cao2012Science,Joseph2015PRL}
and calculated by strong-coupling pair fluctuation theory \citep{Enss2011AoP},
and the thermal conductivity has been considered \citep{Braby2010PRA,Frank2020PRResearch}.
As summarized by two of the present authors in a recent theoretical
analysis \citep{Hu2018PRA}, these developments are promising towards
the full characterization of the second sound propagation in uniform
traps.

In this year (2022), such a dream came true. This feat was accomplished
by X. Li and his colleagues at the University of Science and Technology
of China (USTC) \citep{Li2022Science}, by creating a uniform unitary
Fermi gas of lithium atoms with a fairly large Fermi energy and by
applying a novel high-resolution Bragg spectroscopy. The superfluid
density has been directly measured, with better accuracy. All the
transport coefficients are also successfully determined.

It is fairly non-trivial to take nearly twenty years to fully characterize
the second sound in a strongly interacting unitary Fermi gas. Interestingly,
this is not the end of the journey. The measurements at USTC reveal
an impressive sudden rise in the second sound attenuation and thermal
conductivity near the superfluid phase transition, which can be viewed
as a precursor of critical divergence anticipated near quantum criticality.
Future exploration of the quantum critical region with improved temperature
controllability may open a new era for studying universal critical
dynamics in the strongly interacting regime.

The technique of a uniform box-trap potential also enables the realization
of homogeneous interacting Bose-Einstein condensate (BEC) in two and
three dimensions. In both cases, the propagation of second sound was
observed \citep{Christodoulou2021Nature,Hilker2021arXiv}. The second
sound attenuation however may hardly be accurately determined, due
to the atom loss at the relatively large $s$-wave scattering length,
which is necessary to bring the system into the hydrodynamic regime.

The rest of the short review is organized as follows. In the next
chapter (Sec. II), we present Landau's two-fluid hydrodynamic theory,
and briefly discuss the key features of second sound in strongly interacting
superfluid helium, particularly near the quantum critical regime.
In Sec. III, we then introduce the earlier theoretical treatments
of the second sound of a unitary Fermi gas in harmonic trap, based
on the variational reformulation of the two-fluid hydrodynamic theory.
In Sec. IV and Sec. V, we discuss respectively the two milestone experiments
on the observation of the second sound propagation and on the measurement
of the second sound attenuation in the inhomogeneous and homogeneous
unitary Fermi gases, respectively. In Sec. VI, we summarize the second
sound measurements in weakly interacting Bose gases. Finally, Sec.
VII is devoted to the conclusions and outlooks for some open questions.

\begin{figure}
\centering{}\includegraphics[width=1\columnwidth]{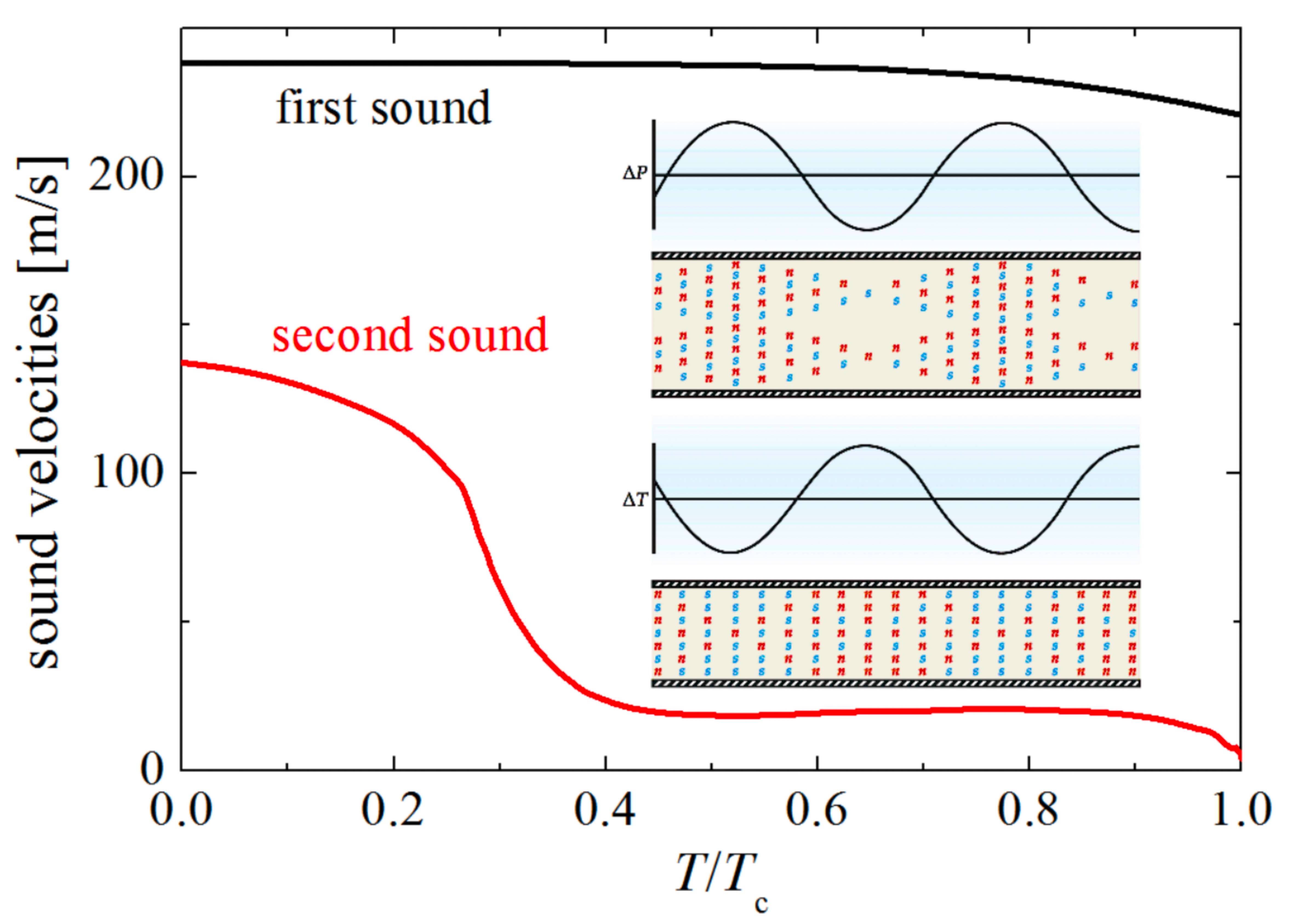}\caption{\label{fig1_HeliumSoundVelocity} First and sound velocities of superfluid
helium as a function of the reduced temperature $T/T_{c}$, where
$T_{c}\simeq2.17$ K is the critical temperature. The inset illustrates
the characteristics of the two sounds: the in-phase motion (first
sound) and the out-of-phase motion (second sound) between the normal
and superfluid components, which are shown by the red circles with
``n'' and the blue circles with ``s''. Adapted from Ref. \citep{Donnelly2009PhysicsToday}
and Ref. \citep{Taylor2009PRA}.}
\end{figure}

\section{Second sound in superfluid helium}

We start by discussing the second sound in superfluid helium. The
existence of a peculiar second sound was predicted by Laszlo Tisza
\citep{Tisza1938Nature}, who first presented the two-fluid hydrodynamic
theory of superfluid helium in 1938. Tisza pictured the superfluid
helium as having two miscible fluids, the superfluid and normal fluid,
which have independent density ($n_{s}$ and $n_{n}$), and move with
different velocities ($\mathbf{v}_{s}$ and $\mathbf{v}_{n}$) and
without mutual interactions. The superfluid does not carry entropy
and flows without dissipation, while the normal fluid carries all
the entropy and does have a finite viscosity $\eta$. The total density
$n$ of superfluid helium is given by the sum of the superfluid density
and the normal-fluid density: $n=n_{s}+n_{n}$. 

A direct consequence of the two interpenetrating fluids is that we
may have two different sound waves, as illustrated in Fig. \ref{fig1_HeliumSoundVelocity}.
First sound is an ordinary sound wave consisting of fluctuations in
total density $n$, in which the superfluid and normal fluid are locked
together to undergo an in-phase motion. It depends very weakly on
the temperature. In contrast, in another sound wave the superfluid
and normal fluid take an out-of-phase motion against each other. As
the total density remains constant to first order, the out-of-phase
motion leads to fluctuations in relative density and hence in entropy.
This creates the entropy wave or temperature wave, which was named
as second sound by Lev Landau. 

The prediction of second sound is a testbed for the two-fluid hydrodynamic
theory. The measurement of second sound velocity turns out to be crucial,
in order to correctly understand how the normal-fluid component forms.
In 1944, second sound of superfluid helium was successfully detected
by Vasilii Peshkov using a standing-wave technique \citep{Peshkov1944}.
To fully account for the observed temperature dependence of the second
sound velocity for temperatures down to $0.4$ K (see the red curve
in Fig. \ref{fig1_HeliumSoundVelocity}), we could apply the celebrated
two-fluid hydrodynamic equations that were extensively developed by
Lev Landau in 1941 \citep{Landau1941PR,Khalatnikov2000Book}:

\begin{align}
m\partial_{t}n+\mathbf{\nabla}\cdot\mathbf{j} & =0,\label{eq:densityHD}\\
m\partial_{t}\mathbf{v}_{s}+\mathbf{\nabla}\left(\mu+V_{ext}\right) & =0,\label{eq:vsHD}\\
\partial_{t}j_{i}+\partial_{i}P+n\partial_{i}V_{ext} & =\partial_{t}\left(\eta\Gamma_{ik}\right),\label{eq:currentHD}\\
\partial_{t}s+\mathbf{\nabla}\cdot\left(s\mathbf{v}_{n}\right) & =\mathbf{\nabla}\cdot\left(\kappa\mathbf{\nabla}T/T\right),\label{eq:entropyHD}
\end{align}
where $\mathbf{j}=m(n_{s}\mathbf{v}_{s}+n_{n}\mathbf{v}_{n})$ is
the current density, $\mu$ is the chemical potential, $P$ is the
pressure, and $s$ is the entropy density. The terms on the right-hand-side
of equations stand for dissipation: $\eta$ and $\kappa$ are the
shear viscosity and thermal conductivity, respectively, and $\Gamma_{ik}\equiv(\partial_{k}v_{ni}+\partial_{i}v_{nk}-2\delta_{ik}\partial_{j}v_{nj}/3)$.
For simplicity, we have omitted various second viscosity terms. We
have also kept only linear terms in the velocities, which is appropriate
for the small-amplitude dynamics such as sound waves. We can then
neglect the relative velocity dependence of the thermodynamic quantities
and use the standard Gibbs-Duhem relation, i.e., $nd\mu=-sdT+dP$.
An external trapping potential $V_{ext}$ is introduced for later
convenience. For superfluid helium, it can be understood as the potential
from the container wall and we can set $V_{ext}=0$ by considering
the thermodynamic limit.

\subsection{Second sound velocity}

In this case, we can look for the small-amplitude solutions that vary
in space and time like plane-waves (i.e., $\delta n,\delta T\propto e^{i(\mathbf{k\cdot}\mathbf{r}-\omega t)}$),
and therefore expand all the thermodynamic quantities around their
equilibrium values. After some lengthy but straightforward manipulation,
we obtain the equation for the sound velocities in the absence of
dissipation \citep{Khalatnikov2000Book},
\begin{equation}
c^{4}-\left(v_{S}^{2}+v^{2}\right)c^{2}+v_{T}^{2}v^{2}=0,\label{eq:velocityEQ}
\end{equation}
where $v_{S}=\sqrt{(\partial P/\partial n)_{\bar{s}}/m}$ and $v_{T}=\sqrt{(\partial P/\partial n)_{T}/m}$
are respectively the adiabatic and isothermal sound velocities, and
we have also defined,
\begin{equation}
v\equiv\sqrt{\frac{n_{s}}{n_{n}}\frac{T\bar{s}^{2}}{m\bar{c}_{V}}},
\end{equation}
with the entropy per particle $\bar{s}\equiv s/n$ and the specific
heat per particle at constant volume $\bar{c}_{V}=T(\partial\bar{s}/\partial T)_{T}$.
By using the thermodynamic relation, $v_{S}^{2}/v_{T}^{2}=c_{P}/c_{V}$,
where $c_{P}$ is the specific heat per volume at constant pressure,
it is convenient to introduce the so-called Landau-Placzek (LP) parameter
\begin{equation}
\epsilon_{\textrm{LP}}\equiv\frac{c_{P}}{c_{V}}-1,
\end{equation}
which quantifies the coupling between the first and second sound.
For superfluid helium, the LP parameter is typically very small, unless
very close to the superfluid lambda transition. As a result, the first
sound velocity can be well approximated by $c_{1}\simeq v_{S}$, and
the second sound velocity $c_{2}\simeq v$. The small LP parameter
indicates that one can hardly excite the second sound by simply modulating
density to set up pressure variations. Indeed, in Peshkov experiment
\citep{Peshkov1944}, a periodic heating was used following the analysis
by Evgeny Lifshitz. This unambiguously established second sound as
a temperature wave. 

The measured second sound velocity can be well explained by using
Landau's picture for normal fluid \citep{Landau1941PR,Khalatnikov2000Book},
i.e., it consists of quasiparticles such as phonons and rotons. In
particular, the rise of second sound velocity below $T<0.9$ K (or
$T<0.4T_{c}$) should be contributed to the collective excitations
of phonons. At sufficiently low temperature, the second sound velocity
then would saturate to $c_{2}=c_{1}/\sqrt{3}$ \citep{Khalatnikov2000Book}.
This anticipation, however, can not be examined experimentally (see
the red dotted curve in Fig. \ref{fig1_HeliumSoundVelocity}), because
the density of normal fluid becomes so low that local thermodynamic
equilibrium cannot be established and the system goes into the collisionless
regime.

\subsection{Second sound attenuation}

\begin{figure}
\centering{}\includegraphics[width=1\columnwidth]{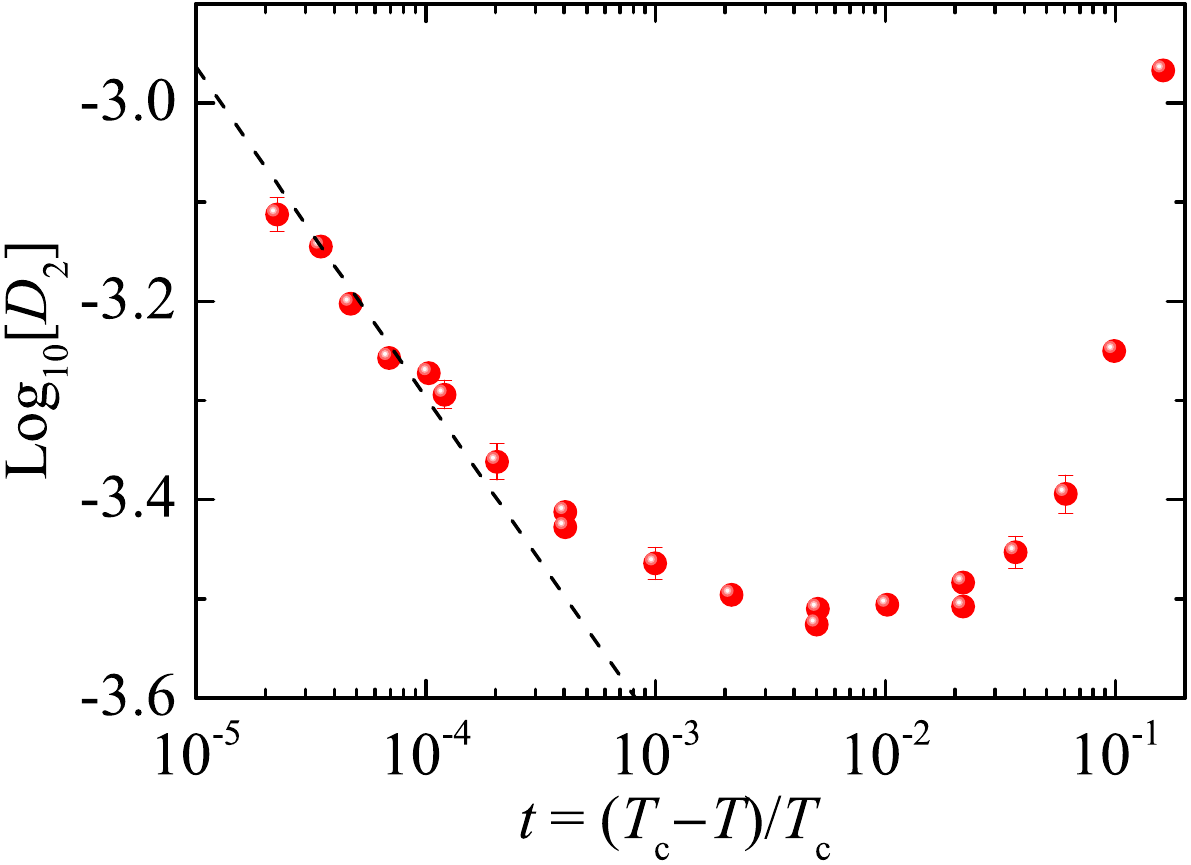}\caption{\label{fig2_Helium2ndSoundAttenuation} Second sound diffusivity $D_{2}$,
in units of cm$^{2}$/s, as a function of the reduced temperature
$t=(T_{c}-T)/T_{c}$. The experimental data (circles) are from Mehrotra
and Ahlers in 1983, measured from the resonance linewidth of cw cavities
\citep{Mehrotra1983PRL}. The dashed line illustrates the critical
behavior $D_{2}\propto t^{-\nu/2}$ near the superfluid lambda transition,
with a critical exponent $\nu\simeq2/3$. }
\end{figure}

The two-fluid hydrodynamic equations can also be solved for sound
waves in the presence of the dissipation terms, where the sound frequency
$\omega$ becomes complex number, indicating sound damping or attenuation.
For second sound, the sound attenuation coefficient $\alpha_{2}\equiv\textrm{Im}$$(\omega/c_{2})$
or the sound diffusivity $D_{2}=2\alpha c_{2}^{3}$/$\omega^{2}$
is given by \citep{Khalatnikov2000Book},
\begin{equation}
D_{2}\simeq\frac{n_{s}}{n_{n}}\left[\frac{4\eta}{3\rho}+\left(\frac{\zeta_{2}}{\rho}+\rho\zeta_{3}-2\zeta_{1}\right)\right]+\frac{\kappa}{\rho c_{P}},
\end{equation}
where $\rho\equiv mn$ is the total mass density and we have included
the contributions of second viscosities $\zeta_{i}$ ($i=1,2,3$).
Near the superfluid lambda transition, the second sound attenuation
is dominated by the thermal damping term $D_{T}\equiv\kappa/(\rho c_{P})$.
In Fig. \ref{fig2_Helium2ndSoundAttenuation}, we show the experimental
data of the second sound diffusivity, reported by Mehrotra and Ahlers
in 1983 \citep{Mehrotra1983PRL}. 

Historically, the measurement of second sound attenuation in superfluid
helium plays a crucial role to establish the celebrated dynamic scaling
theory of superfluid phase transition \citep{Ferrell1967PRL,Hohenberg1977RMP}.
By extending the static scaling ideas that involves the coherent length
$\xi(T)\propto(T_{c}-T)^{-\nu}$, where $\nu\simeq2/3$ is the critical
exponent, R. A. Ferrell and his co-workers stated that the dynamics
near the phase transition is governed by a characteristic frequency
$\omega^{*}=c_{2}/\xi$. As the second sound velocity vanishes like
$c_{2}\propto(T_{c}-T)^{\nu/2}$ close to the transition, both the
sound diffusivity $D_{2}$ and the thermal conductivity $\kappa$
then would diverge as $\left|T_{c}-T\right|$$^{-\nu/2}.$ This critical
divergence can be clearly seen in Fig. \ref{fig2_Helium2ndSoundAttenuation},
where the critical scaling law $\left|t\right|^{-\nu/2}$ is illustrated
by the dashed line. The data seems to deviate from the critical divergence
at the reduced temperature $t\gtrsim5\times10^{-4}$, strongly indicating
the existence of a quantum critical regime at $\left|t\right|<5\times10^{-4}$
or $\left|T_{c}-T\right|<1$ mK.

It is worth noting that a minimum second sound diffusivity $(D_{2})_{\textrm{min}}\simeq10^{-3.54}$
cm$^{2}$/s appears at $\left|t\right|\sim10^{-2}$ \citep{Donnelly1998}.
In terms of $\hbar/m\simeq1.59\times10^{-4}$ cm$^{2}$/s for the
mass of helium atoms, we find that 
\begin{equation}
\left(D_{2}\right)_{\textrm{min}}\simeq1.8\frac{\hbar}{m}.\label{eq:D2minHeII}
\end{equation}
A similar minimum occurs in the thermal diffusivity $D_{T}$ at the
same temperature range $\left|t\right|\sim10^{-2}$ just above the
superfluid transition \citep{Donnelly1998}: $(D_{T})_{\textrm{min}}\simeq1.5\hbar/m$.

\subsection{Hydrodynamic regime vs quantum criticality}

Landau's two-fluid hydrodynamic equations are applicable in the limits
of long wavelength and low frequency \citep{Khalatnikov2000Book},
i.e., $k\xi\ll1$ and $\omega\tau\ll1$, where $\tau$ is a characteristic
collision time. 

Very close to the superfluid lambda transition, the former condition
$k\xi\ll1$ necessarily breaks down, due to the divergent correlation
length $\xi\rightarrow\infty$ \citep{Hohenberg1977RMP}. A reasonable
estimate of $\xi$ in the superfluid phase is given by \citep{Hohenberg1977RMP},
$\xi\sim r_{0}\left|t\right|^{-\nu}$, where $r_{0}\simeq2.22$ Å
is the mean-distance between two helium atoms. This seems to impose
a stringent restriction for the use of neutron scattering, since the
transferred momentum $k$ in the neutron-scattering experiment is
typically larger than $0.1$ Å$^{-1}$ \citep{Griffin1993Book} and
hence the hydrodynamic regime is difficult to reach. Brillouin light
scattering can instead be used to measure the density correlation
function (i.e., dynamical structure factor) in the hydrodynamic regime
\citep{Tarvin1977PRB}, with a momentum $kr_{0}\sim0.004$ or $k\xi\simeq0.004\left|t\right|^{-\nu}$.
However, one may hardly reach the quantum critical regime at $\left|t\right|<5\times10^{-4}$,
as suggested by the data in Fig. \ref{fig2_Helium2ndSoundAttenuation}.
This is consistent with the observation in the Brillouin light scattering
experiment \citep{Tarvin1977PRB} that, below the superfluid transition
the second sound attenuation starts to saturate at $\left|T_{c}-T\right|<1$
mK.

On the other hand, the latter hydrodynamic condition $\omega\tau\ll1$
becomes more difficult to satisfy at very low temperature. In the
measurement of first sound attenuation, one typically observes a maximum
at $T\sim0.4T_{c}\simeq0.9$ K \citep{Woods1973RPP}, due to the transition
from the hydrodynamic regime to the collisionless regime.

\section{Second sound in unitary Fermi gas: Earlier theoretical treatments}

Let us now look back to year 2005. At that time, Fermi superfluidity
near a Feshbach resonance has been unambiguously verified, by measuring
nonzero condensate fraction \citep{Regal2004PRL,Zwierlein2004PRL}
and by observing vortex lattice \citep{Zwierlein2005Nature}. Yet,
there was no trace for the most important second sound, presumably
due to two technical difficulties. On the one hand, the thermodynamic
functions or equations of state of a unitary Fermi gas were less known.
In particular, the superfluid density had only been calculated using
the mean-field Bardeen-Cooper-Schrieffer (BCS) theory, which is not
so accurate. On the other hand, a unitary Fermi gas had have to be
confined in a harmonic trap. While zero-temperature first sound such
as the breathing mode of a trapped unitary Fermi gas was theoretically
discussed \citep{Stringari2004EPL,Hu2004PRL,Astrakharchik2005PRL}
and experimentally explored in detail \citep{Kinast2004PRL,Bartenstein2004PRL,Altmeyer2007PRL},
at finite temperature both first and second sound modes in a harmonic
trap were not considered.

\begin{figure}
\centering{}\includegraphics[width=1\columnwidth]{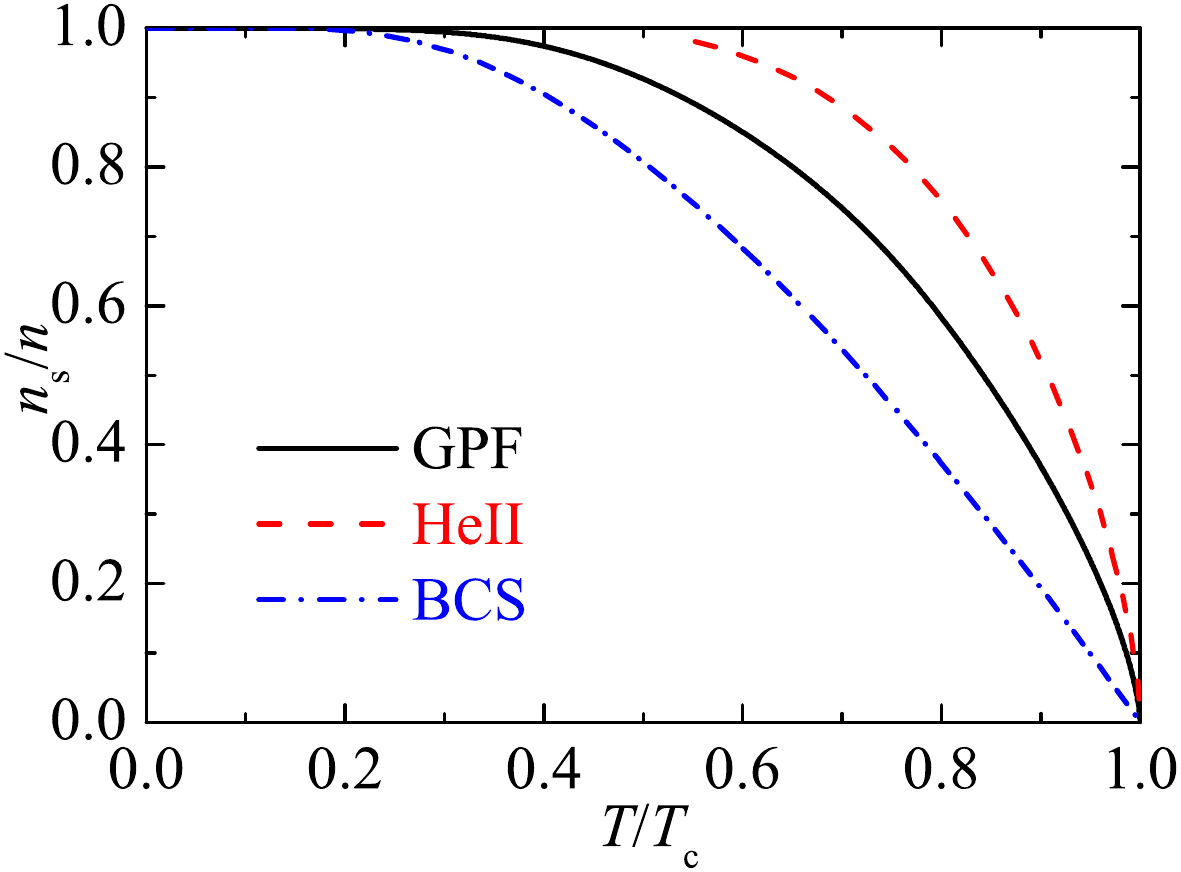}\caption{\label{fig3_ns} The superfluid fraction of a unitary Fermi gas, predicted
by the Gaussian pair fluctuation (GPF) theory (black solid line) and
mean-field BCS theory (blue dot-dashed line). For comparison, the
superfluid fraction of superfluid helium is also shown by a red dashed
line. Note that, the GPF theory predicts a spurious bend-back structure
near the superfluid transition \citep{Fukushima2007PRA}. This unphysical
behavior has been rectified \citep{Taylor2008PRA}, by extrapolating
the reliable results below $T_{c}$ with the correct critical scaling
behavior $n_{s}/n\propto(1-T/T_{c})^{-\nu}$, with $\nu\simeq2/3$.
Here, the critical temperature of a unitary Fermi gas is $T_{c}\simeq0.17T_{F}$,
as determined experimentally. The Fermi temperature $T_{F}=\varepsilon_{F}/k_{B}=\hbar^{2}k_{F}^{2}/(2mk_{B})$
and the Fermi wavevector $k_{F}=(3\pi^{2}n)^{1/3}$ at density $n$.}
\end{figure}

The first problem was partly solved with the development of strong-coupling
diagrammatic theories at the BEC-BCS crossover \citep{Hu2006EPL,He2015PRA}.
In particular, the superfluid density has been calculated by the Gaussian
pair fluctuation (GPF) theory \citep{Fukushima2007PRA}, in which
strong bosonic pair-fluctuations at the Gaussian level are taken into
account on top of the standard BCS mean-field theory (see also the
simplified treatment of the bosonic degree of freedom in Ref. \citep{Salasnich2010PRA}
and Ref. \citep{Baym2013PRA}). As shown in Fig. \ref{fig3_ns}, the
GPF theory predicts a larger superfluid density than the BCS mean-field
theory. Compared with superfluid helium, unitary Fermi gas turns out
to have a smaller superfluid fraction. It is worth noting that, for
the superfluid density of a unitary Fermi gas the GPF result remains
as the best theoretical prediction so far \citep{Taylor2008PRA},
since \emph{ab-initio} quantum Monto Carlo simulation is still not
available towards the meaningful thermodynamic limit.

\begin{figure}
\centering{}\includegraphics[width=1\columnwidth]{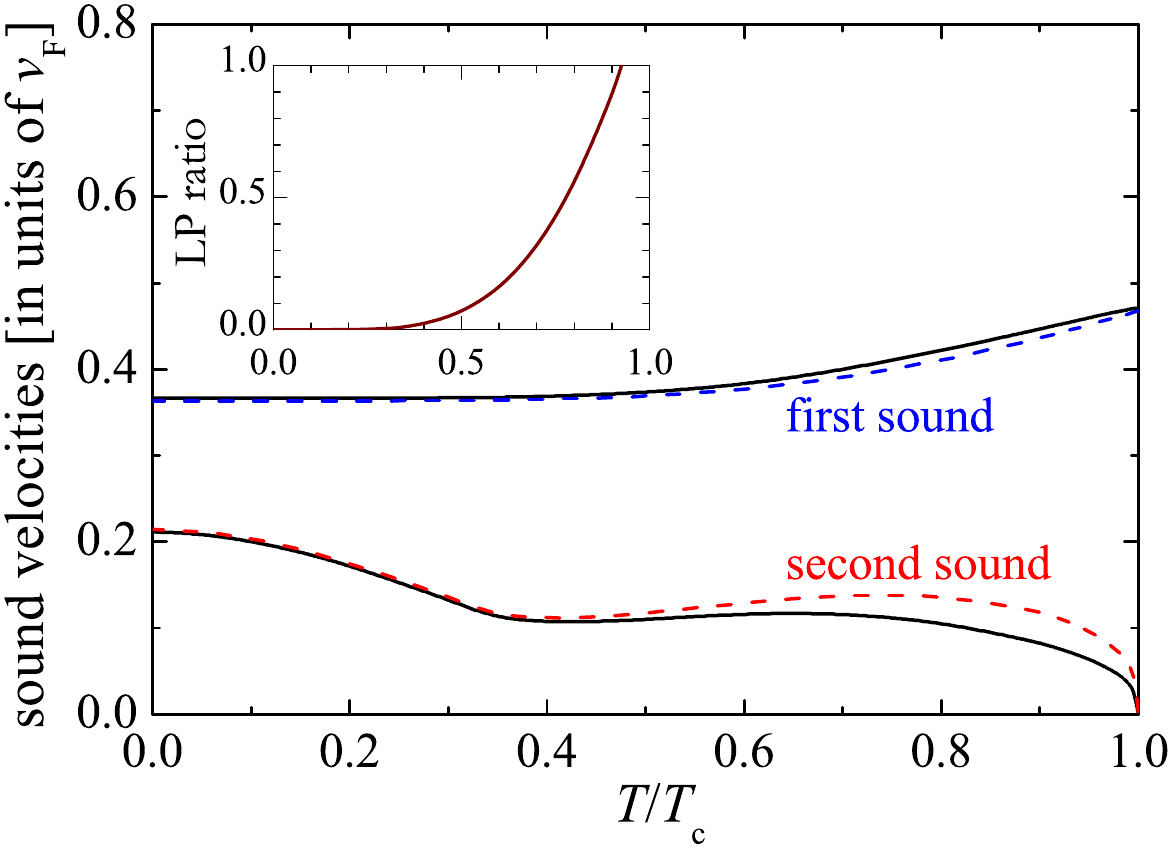}\caption{\label{fig4_SoundVelocity} First and second sound velocities of a
uniform Fermi gas at unitarity (solid lines), predicted by the GPF
theory. The units of the velocity is $v_{F}=\hbar k_{F}/m$. The uncoupled
first and second sound velocities given by $v_{S}$ and $v$ are shown
as dashed and dotted lines, respectively. The inset shows the Landau-Placzek
ratio as a function of $T/T_{c}$. From Ref. \citep{Taylor2009PRA}.}
\end{figure}

The GPF theory also predicts reasonably accurate thermodynamic functions
for unitary Fermi gas \citep{Hu2006EPL,Hu2007NaturePhysics}. By applying
the dissipationless two-fluid hydrodynamic equations, it is then straightforward
to calculate the first and second sound velocities in the unitary
limit \citep{Taylor2009PRA}, as reported in Fig. \ref{fig4_SoundVelocity}.
Remarkably, the temperature dependence of second sound velocity in
unitary Fermi gas is qualitatively similar to that of superfluid helium,
in spite of the entirely different statistics of the two systems.
At low temperatures, we may attribute the similarity to gapless phonon
excitations, which are the dominant quasiparticles of both systems.
At temperatures close to the superfluid transition, fermionic gapped
single-particle excitations become important in unitary Fermi gas.
It seems reasonable to argue that these single-particle excitations
may play the same role as rotons in superfluid helium, which are also
gapped, single-particle-like excitations.

Despite the similar temperature dependence in second sound velocity,
there is an important difference between unitary Fermi gas and superfluid
helium. As shown in the inset of Fig. \ref{fig4_SoundVelocity}, unlike
superfluid helium the LP ratio of a unitary Fermi gas can be very
significant already at $T\sim0.6T_{c}$, suggesting a strong coupling
between first and second sound. As a result, we should be able to
observe the temperature wave of second sound from the density response.
This solves a potential detection problem, since we can hardly have
a local thermometer with cold-atoms. The density response, however,
is easy to measure, by using Bragg scattering technique \citep{Stenger1999PRL}.

A significant LP ratio also implies that the second sound velocity
deviates from $v$. By solving Eq. (\ref{eq:velocityEQ}), we find
that the expansion \citep{Hu2010NJP},
\begin{eqnarray}
c_{1}^{2} & = & v_{S}^{2}\left[1+\epsilon_{\textrm{LP}}x+\cdots\right],\label{eq:c1}\\
c_{2}^{2} & = & \frac{c_{V}}{c_{P}}v^{2}\left[1-\epsilon_{\textrm{LP}}x+\cdots\right],\label{eq:c2}
\end{eqnarray}
in terms of the parameter $x\equiv(v^{2}c_{V})/(v_{S}^{2}c_{P})$,
which is very small in the temperature region of interest (i.e., $T>0.6T_{c}$).
Thus, the second sound velocity can be well approximated by,
\begin{equation}
c_{2}=\frac{v}{\sqrt{c_{P}/c_{V}}}=\sqrt{\frac{n_{s}}{n_{n}}\frac{T\bar{s}^{2}}{m\bar{c}_{P}}}.\label{eq:c2Approximate}
\end{equation}

\subsection{Second sound in a harmonic trap}

The second problem related to the harmonic trapping potential $V_{ext}(\mathbf{r})$
was solved by Allan Griffin and his colleagues, with a reformulation
of Landau's dissipationless two-fluid hydrodynamic equations in a
variational form \citep{Taylor2005PRA,Taylor2008PRA,Taylor2009PRA}.
This reformulation is crucial, since the brute-force solution of Landau's
two-fluid equations turns out to be difficult \citep{Shenoy1998PRL,He2007PRA}. 

By introducing the displacement fields $\mathbf{u}_{s}$ and $\mathbf{u}_{n}$
through $\mathbf{v}_{s}(\mathbf{r},t)=\partial\mathbf{u}_{s}(\mathbf{r},t)/\partial t$
and $\mathbf{v}_{n}(\mathbf{r},t)=\partial\mathbf{u}_{n}(\mathbf{r},t)/\partial t$,
they found that the solution of hydrodynamic modes with frequency
$\omega$ (i.e., $\mathbf{u}_{s}(\mathbf{r},t)=\mathbf{u}_{s}(\mathbf{r})e^{-i\omega t}$
and $\mathbf{u}_{n}(\mathbf{r},t)=\mathbf{u}_{n}(\mathbf{r})e^{-i\omega t}$
) can be derived by minimizing the variational expression \citep{Taylor2005PRA},
\begin{eqnarray}
\mathcal{S} & = & \frac{1}{2}\int d\mathbf{r}\left[m\omega^{2}\left(n_{s}\mathbf{u}_{s}^{2}+n_{n}\mathbf{u}_{n}^{2}\right)-\left(\frac{\partial\mu}{\partial n}\right)_{s}\left(\delta n\right)^{2}\right.\nonumber \\
 &  & \left.-2\left(\frac{\partial T}{\partial n}\right)_{s}\delta n\delta s-\left(\frac{\partial T}{\partial s}\right)_{n}\left(\delta s\right)^{2}\right],\label{eq:usunAction}
\end{eqnarray}
where the superfluid density $n_{s}(\mathbf{r})$, the normal density
$n_{n}(\mathbf{r})$, and the various equilibrium thermodynamic variables
$(\partial\mu/\partial n)_{s}$, $(\partial T/\partial n)_{s}$ and
$(\partial T/\partial s)_{n}$ are position dependent due to the harmonic
trap $V_{ext}(\mathbf{r})$, and their spatial dependence can be calculated
within the standard local density approximation. The action Eq. (\ref{eq:usunAction})
includes the kinetic energy terms due to the displacement fields of
the superfluid component ($\mathbf{u}_{s}(\mathbf{r})$) and of the
normal component ($\mathbf{u}_{n}(\mathbf{r})$), and the potential
energy terms arising from the density fluctuation $\delta n$ and
entropy fluctuation $\delta s$, which are respectively given by,
\begin{eqnarray}
\delta n\left(\mathbf{r}\right) & \equiv & -\mathbf{\nabla}\cdot\left(n_{s}\mathbf{u}_{s}+n_{n}\mathbf{u}_{n}\right),\\
\delta s\left(\mathbf{r}\right) & \equiv & -\mathbf{\nabla}\cdot\left(s\mathbf{u}{}_{n}\right).
\end{eqnarray}
The minimization of the action can be easily carried out by applying
a variational polynomial \emph{Ansatz} for the displacement fields
\citep{Taylor2009PRA}.

The variational expression of the two-fluid hydrodynamic theory is
very useful to understand the decoupled first and sound modes in traps.
For the pure in-phase mode of first sound, we can insert $\mathbf{u}_{s}(\mathbf{r})=\mathbf{u}_{n}(\mathbf{r})=\mathbf{u}^{(1)}(\mathbf{r})$
into the action Eq. (\ref{eq:usunAction}) and take the variation
with respect to $\mathbf{u}^{(1)}(\mathbf{r})$. It is straightforward
to show that 
\begin{eqnarray}
m\omega_{1}^{2}\mathbf{u}^{(1)} & = & -\frac{1}{n}\mathbf{\nabla}\left[n\left(\frac{\partial P}{\partial n}\right)_{\bar{s}}\mathbf{\nabla}\cdot\mathbf{u}^{(1)}\right]+\mathbf{\nabla}\left(\mathbf{u}^{(1)}\cdot\mathbf{\nabla}V_{ext}\right)\nonumber \\
 &  & -\mathbf{\left(\mathbf{\nabla}\cdot\mathbf{u}^{(1)}\right)\nabla}V_{ext},\label{eq:FirstsoundTrap}
\end{eqnarray}
which without the harmonic trap ($V_{ext}=0$) gives the plane wave
solution $\propto e^{i\mathbf{k\cdot}\mathbf{r}}$ with dispersion
$\omega_{1}=v_{S}k$. For the pure out-of-phase mode of second sound,
the density fluctuation $\delta n=0$ and we then inset $n_{s}\mathbf{u}_{s}^{(2)}(\mathbf{r})=-n_{n}\mathbf{u}_{n}^{(2)}(\mathbf{r})$
into Eq. (\ref{eq:usunAction}) to obtain, 
\begin{equation}
m\omega_{2}^{2}\mathbf{u}_{s}^{(2)}=-\frac{s}{n}\mathbf{\nabla}\left[\frac{1}{n}\left(\frac{\partial T}{\partial\bar{s}}\right)_{n}\mathbf{\nabla}\cdot\left(\frac{sn_{s}}{n_{n}}\mathbf{u}_{s}^{(2)}\right)\right].\label{eq:SecondsoundTrap}
\end{equation}
In the absence of the trapping potential, it gives rise to the plane
wave solution with dispersion $\omega_{2}=vk$, as anticipated.

\begin{figure}
\begin{centering}
\includegraphics[width=1\columnwidth]{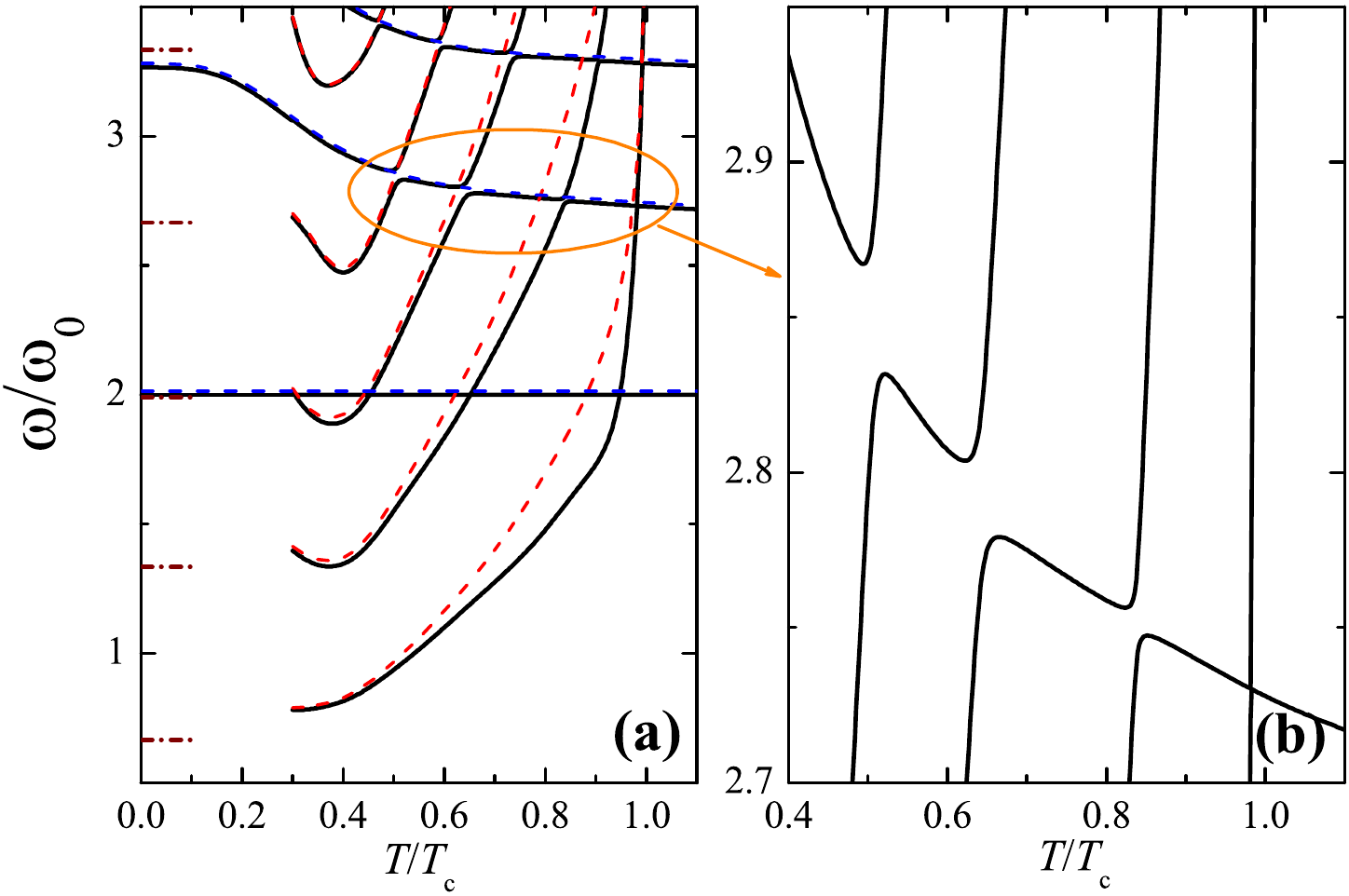}
\par\end{centering}
\begin{centering}
\includegraphics[width=1\columnwidth]{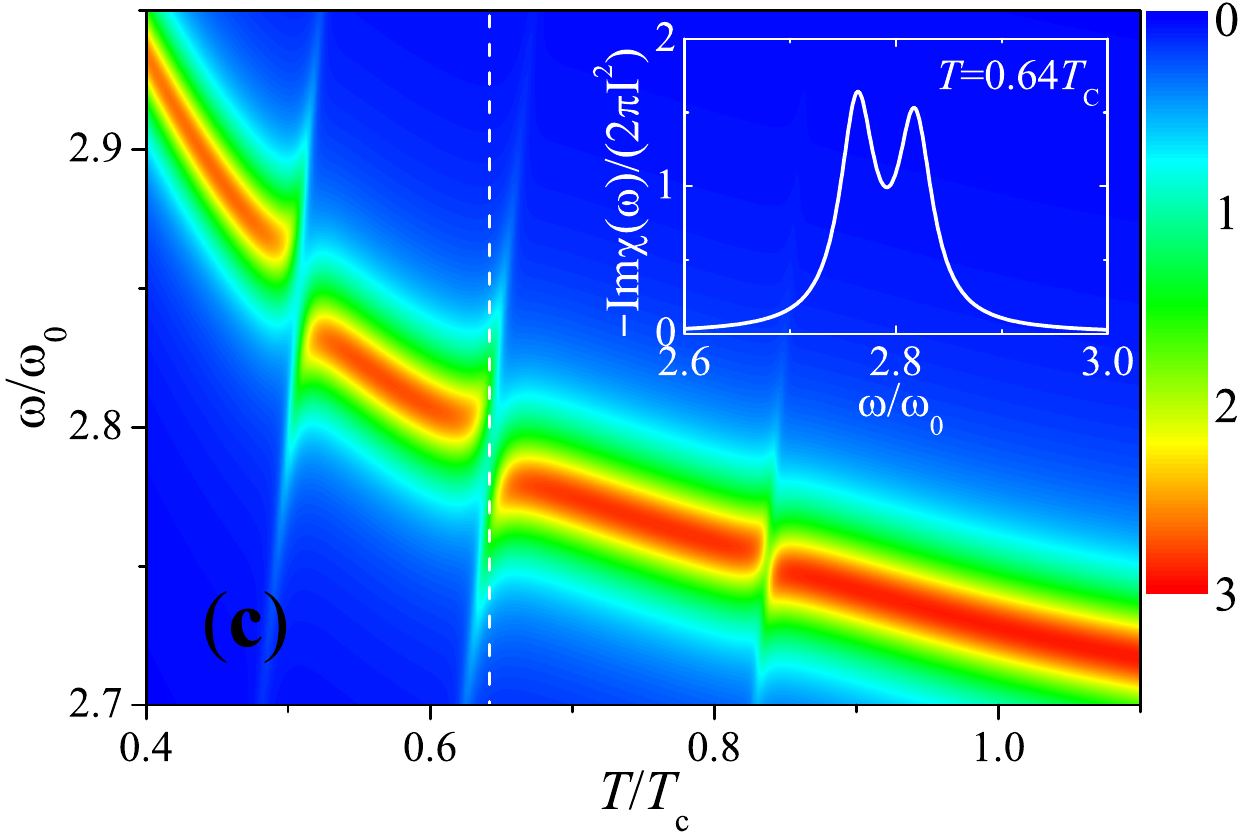}
\par\end{centering}
\centering{}\caption{\label{fig5_TrapDensityResponse} (a) Two-fluid modes of a unitary
Fermi gas in an isotropic harmonic trap. The full solutions of the
two-fluid equations are given by the solid lines. The first sound
modes (blue dashed lines) given by Eq. (\ref{eq:FirstsoundTrap})
are shown along with the second sound modes (red dotted lines) given
by Eq. (\ref{eq:SecondsoundTrap}). The dashed lines at $T=0$ show
the analytic results for the second sound frequency. (b) A blow-up
of part of the left panel showing the hybridization between second
sound and the $n=1$ first sound mode. (c) The two-dimensional contour
plot of the density response $\textrm{Im}\chi(\omega)$ in the vicinity
of the $n=1$ first sound mode. The delta functions in $\textrm{Im}\chi(\omega)$
have been broadened by $0.02\omega_{0}$ for visibility. The inset
shows a double peak structure at the hybridization point $T\simeq0.64T_{c}$.
Adapted from Ref. \citep{Taylor2009PRA}.}
\end{figure}

The coupled first sound and second sound in an isotropic harmonic
trap $V_{ext}(r)=m\omega_{0}^{2}r^{2}/2$ have been solved for the
compressional modes with angular momentum $l=0$, by applying the
following variational \emph{Ansatz} \citep{Taylor2009PRA}, 
\begin{equation}
u_{s(n)}=r\sum_{j}^{N-1}a_{s(n),j}r^{j},
\end{equation}
with $N$ variational parameters $\{a_{s,j},a_{n,j}\}$. Figure \ref{fig5_TrapDensityResponse}(a)
shows the mode frequencies as a function of the temperature, obtained
with $N=8$, which is large enough to have convergent results for
the variation calculations. Also shown are the frequencies of the
decoupled first sound {[}$\omega_{1}(n)$, blue dashed lines{]} and
second sound {[}$\omega_{2}(n)$, red dotted lines{]} modes given
by Eqs. (\ref{eq:FirstsoundTrap}) and (\ref{eq:SecondsoundTrap}),
respectively, where $n=0,1,2,\cdots$ is the number of radial nodes.
First sound appears to be the ``horizontal'' branches, with mode
frequency decreases with increasing temperature. In contrast, at $T>0.4T_{c}$
the mode frequency of second sound increases rapidly with temperature
and becomes divergent close to the superfluid transition, i.e., $\omega_{2}\propto(1-T/T_{c})^{(\nu-1)/2}$,
where $\nu\simeq2/3$ is the critical exponent. This peculiar divergence
is found to be related to the critical behavior of superfluid density
\citep{Taylor2009PRA}.

To experimentally probe the second sound branches, the authors proposed
to use Bragg scattering to measure the density response \citep{Taylor2009PRA},
thanks to the significant LP ratio of unitary Fermi gas and hence
the strong coupling between first second and second sound. This is
evident in Fig. \ref{fig5_TrapDensityResponse}(b) from the avoided
crossings and hybridizations, which mean that second sound will manifest
itself in the density response at the crossing points. By adding a
density perturbation of the form $\delta V\propto f(\mathbf{r})e^{-i\omega t}$
to the variational action Eq. (\ref{eq:usunAction}), it is easy to
derive the density response function $\chi(\omega)=\int d\mathbf{r}f(\mathbf{r})\delta n(\mathbf{r})$,
with imaginary part ($Z_{1}+Z_{2}=1$)
\begin{equation}
-\textrm{Im}\chi\left(\omega\right)\propto Z_{1}\delta\left(\omega^{2}-\tilde{\omega}_{1}^{2}\right)+Z_{2}\delta\left(\omega^{2}-\tilde{\omega}_{2}^{2}\right),
\end{equation}
near a crossing point with mode frequencies $\tilde{\omega}_{1}$
and $\tilde{\omega}_{2}$. The weights $Z_{1}$ and $Z_{2}$ becomes
equal at the crossing point. Fig. \ref{fig5_TrapDensityResponse}(c)
clearly shows the bimodal structure in the vicinity of hybridization
between the $n=1$ first sound mode and second sound. The inset highlights
the hybridization point at $T\simeq0.64T_{c}$, where the first sound
and second sound contribute equally to the density response (i.e.,
$Z_{1}\sim Z_{2}$). The frequency splitting $\Delta\omega\sim0.06\omega_{0}$
between the two peaks is large enough to be resolved in experiments.
On the other hand, for the low-lying collective modes in harmonic
traps, both viscous damping and thermal damping are supposed to be
small, so the double peak structure shown in the inset could be robust.

\begin{figure}
\begin{centering}
\includegraphics[width=1\columnwidth]{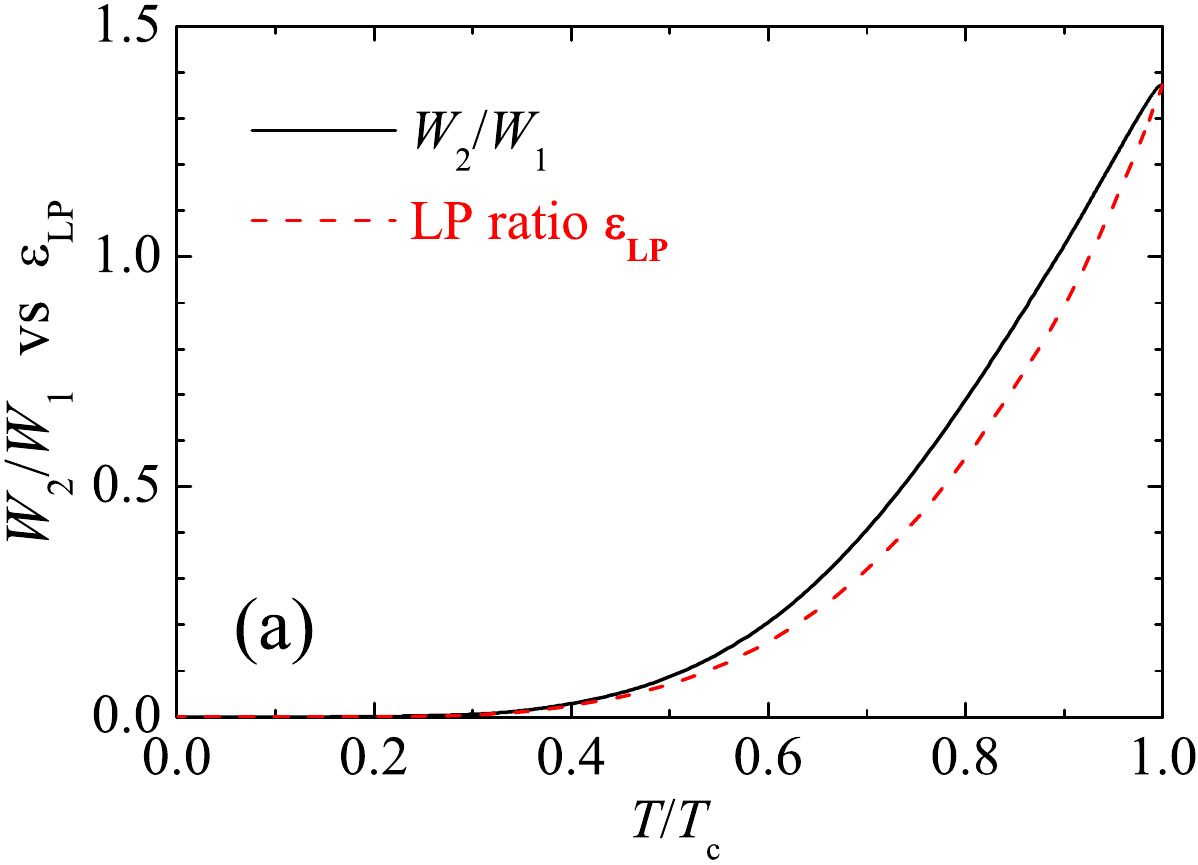}
\par\end{centering}
\begin{centering}
\includegraphics[width=1\columnwidth]{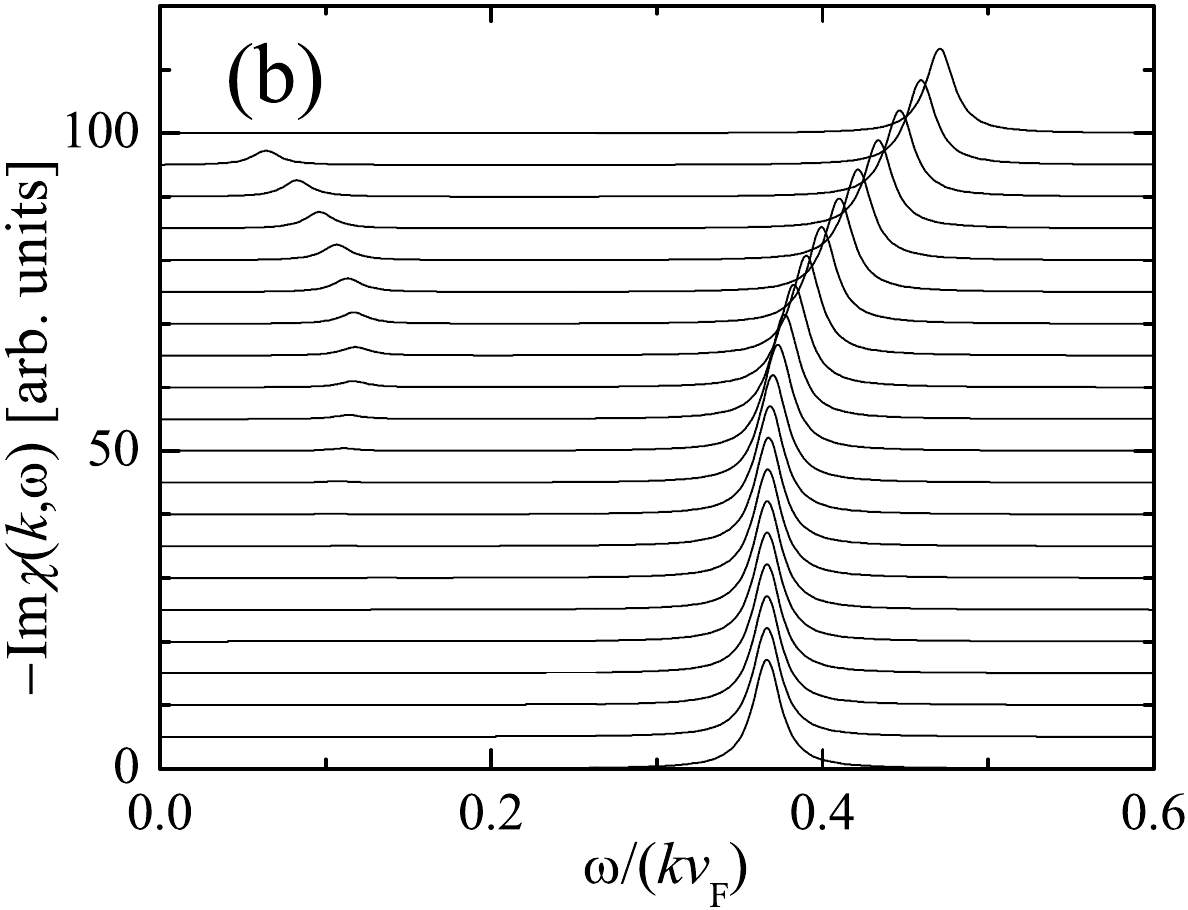}
\par\end{centering}
\centering{}\caption{\label{fig6_UniformDensityResponse} (a) Comparison between the ratio
of the second and first sound amplitudes in the dynamic structure
factor ($W_{2}/W_{1}$) and the Landau--Placzek ratio $\epsilon_{\textrm{LP}}$
in a unitary Fermi gas. (b) The imaginary part of the density response
function at $k=0.1k_{F}$ and at different temperatures (each temperature
is offset). The temperature is increased from zero to $T_{c}$, in
steps of $0.05T_{c}$. For clarity, the delta functions in Eq. (\ref{eq:ImKappaUniform})
are plotted as Lorentzians with a width $0.01kv_{F}$. Adapted from
Ref. \citep{Hu2010NJP}.}
\end{figure}

\subsection{Second sound propagation in a uniform system}

To better understand the second sound of a unitary Fermi gas, Allan
Griffin and his colleagues also discussed in detail the sound propagation
in the uniform configuration \citep{Hu2010NJP}. By setting $V_{ext}=0$
and adding a density perturbation with a given momentum $\mathbf{k}$,
i.e., $\delta V\propto e^{i(\mathbf{k}\cdot\mathbf{r}-\omega t)}$,
to the variation action, it is straightforward to derive the density
response function \citep{Hu2010NJP},
\begin{equation}
\chi\left(k,\omega\right)=\frac{nk^{2}}{m}\frac{\omega^{2}-v^{2}k^{2}}{\left(\omega^{2}-c_{1}^{2}k^{2}\right)\left(\omega^{2}-c_{2}^{2}k^{2}\right)},\label{eq:dissipationlessDensityResponse}
\end{equation}
with imaginary part ($\omega>0$),
\begin{equation}
-\frac{\textrm{Im}\chi}{\pi}=\frac{nk}{2m}\left[\frac{Z_{1}}{c_{1}}\delta\left(\omega-c_{1}k\right)+\frac{Z_{2}}{c_{2}}\delta\left(\omega-c_{2}k\right)\right],\label{eq:ImKappaUniform}
\end{equation}
where $Z_{1}\equiv(c_{1}^{2}-v^{2})/(c_{1}^{2}-c_{2}^{2})$ and $Z_{2}=1-Z_{1}$.
By defining $W_{1}=Z_{1}/c_{1}^{2}$ and $W_{2}=Z_{2}/c_{2}^{2}$,
we find that the relative weight of second sound and first sound in
the Bragg scattering is given by \citep{Hu2010NJP}
\begin{equation}
\frac{Z_{2}/c_{2}}{Z_{1}/c_{1}}=\frac{W_{2}}{W_{1}}\frac{c_{2}}{c_{1}}.
\end{equation}
For unitary Fermi gas, the ratio $W_{2}/W_{1}$ is roughly equal to
the LP ratio $\epsilon_{\textrm{LP}}$, as shown in Fig. \ref{fig6_UniformDensityResponse}(a).
Therefore, although the LP ratio is significant close to the superfluid
transition, the second sound signal in the Bragg scattering measurement
would be still much weaker than the first sound signal, due to the
additional reduction factor $c_{2}/c_{1}$, which is less than $0.3$
in the temperature region of interest. The weak second sound signal
in the density response $-\textrm{Im}\chi$ is explicitly illustrated
in Fig. \ref{fig6_UniformDensityResponse}(b).

\begin{figure}
\begin{centering}
\includegraphics[width=1\columnwidth]{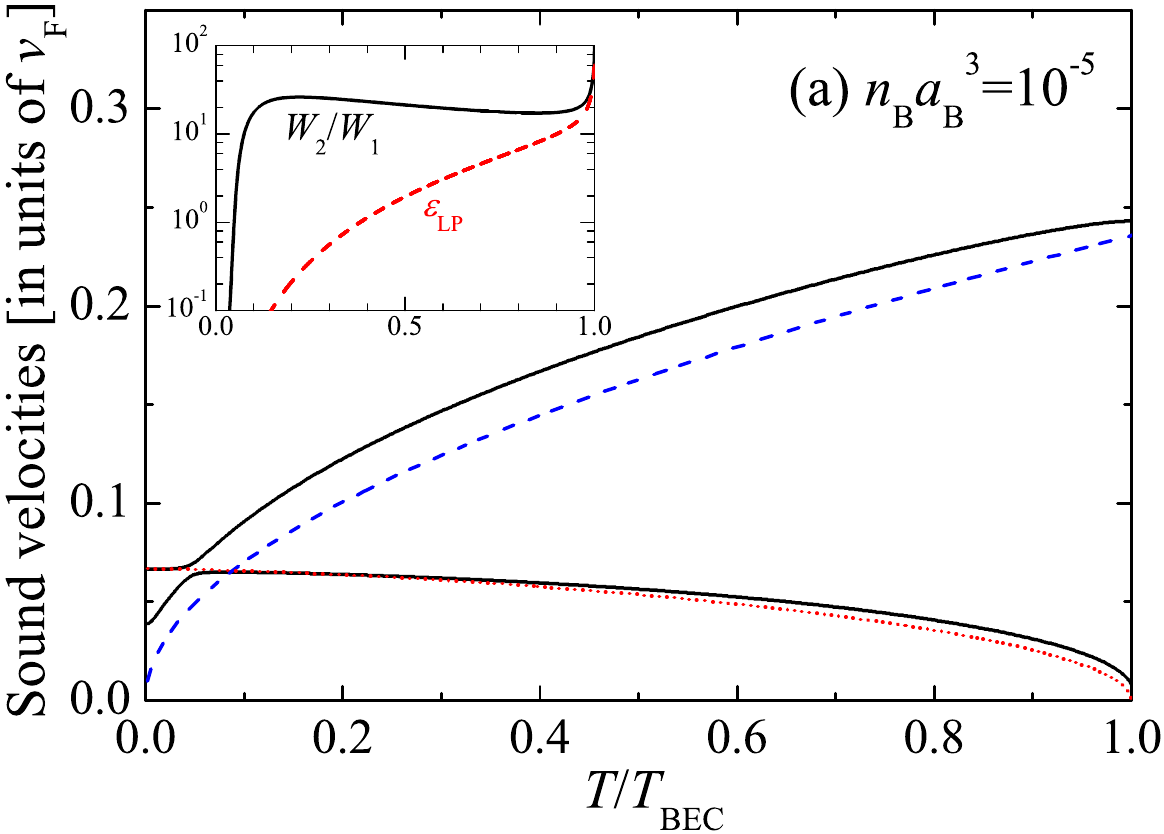}
\par\end{centering}
\begin{centering}
\includegraphics[width=1\columnwidth]{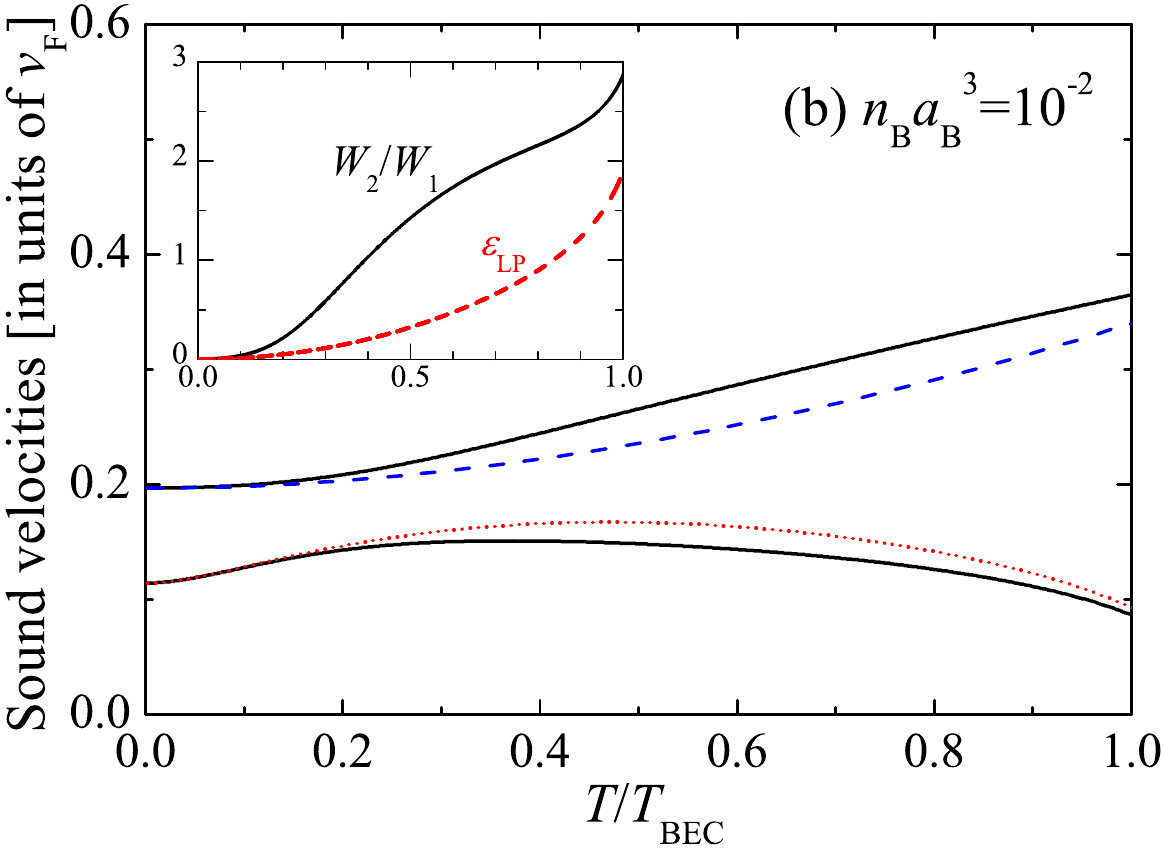}
\par\end{centering}
\centering{}\caption{\label{fig7_BECSoundVelocity} First and second sound velocities in
an interacting molecular Bose gas with gas parameter $n_{B}a_{B}^{3}=10^{-5}$
(a) and $n_{B}a_{B}^{3}=10^{-2}$ (b). In (a), the blue dashed and
red dotted lines show $c_{\textrm{HF}}(T)$ and $c_{B}(T)$; while
in (b) they are, respectively, the approximate velocities $c_{1}=v_{S}$
and $c_{2}=v/\sqrt{c_{P}/c_{V}}$. The two inset show the LP ratio
$\epsilon_{\textrm{LP}}$ (red dashed line) and the weight ratio $W_{2}/W_{1}$
(black solid line). Adapted from Ref. \citep{Hu2010NJP}.}
\end{figure}

\subsection{Unitary Fermi gas vs weakly interacting Bose gas}

The similarity of second sound in unitary Fermi gas and superfluid
helium strongly indicates that the qualitative behavior of second
sound is insensitive to the statistics of underlying particles. Instead,
the inter-particle interaction strength may play an important role.
To confirm such an idea, Allan Griffin and his colleagues investigated
the sound propagation in interacting molecular Bose gas of tightly
bound Cooper pairs of mass $m_{B}=2m$ with tunable interaction strength
\citep{Hu2010NJP}, as shown in Fig. \ref{fig7_BECSoundVelocity}.
In the BEC limit, the molecular scattering length $a_{B}\simeq0.6a$
\citep{Petrov2004PRL} and the density $n_{B}=n/2$. In their calculations,
the thermodynamic functions in Landau's two-fluid hydrodynamic equations
are calculated by using the standard Hartree-Fock-Bogoliubov-Popov
theory at finite temperature and the superfluid density is approximated
by the condensate density, i.e., $n_{s}(T)\simeq n_{c}(T)\leq n_{B}$.

For a weakly interacting Bose gas with gas parameter $na^{3}=10^{-5}$
{[}Fig. \ref{fig7_BECSoundVelocity}(a){]}, first sound and second
sound show an interesting hybridization at very low temperature \citep{Lee1959PR,Griffin1997PRA}.
This hybridization is due to the large compressibility of the weakly
interacting Bose gas, which leads to a large LP ratio $\epsilon_{\textrm{LP}}$
or $W_{2}/W_{1}$, as shown in the inset. It is natural to identify
the upper and lower branches of sound velocities as first sound and
second sound. Below the hybridization point, the first sound velocity
approaches the Bogoliubov phonon velocity \citep{Griffin1997PRA,Hu2010NJP}
\begin{equation}
c_{B}(T)=\frac{\hbar}{m_{B}}\sqrt{4\pi n_{s}(T)a_{B}}
\end{equation}
and first sound corresponds to an oscillation of the condensate component.
Above the hybridization point, the role of first sound and second
sound exchanges. It is second sound that propagates with the phonon
velocity $c_{B}$. Consequently, second sound in dilute molecular
Bose gas (above the hybridization temperature) is essentially an oscillation
of the condensate at all temperatures, with a static thermal cloud.
In contrast, first sound describes an oscillation of the thermal cloud
in the presence of a stationary condensate. In the Hartree-Fock approximation,
its velocity is given by \citep{Griffin1997PRA},
\begin{equation}
c_{\textrm{HF}}(T)=\sqrt{\frac{8\pi\hbar^{2}a_{B}n_{n}(T)}{m_{B}^{2}}+\frac{5k_{B}T}{3m_{B}}\frac{g_{5/2}(1)}{g_{3/2}(1)}},\label{eq:cHF}
\end{equation}
where $g_{n}(z)=\sum_{l=1}^{\infty}z^{l}/l^{n}$ is the usual Bose-Einstein
function and $n_{n}(T)\simeq n-n_{c}(T)$ is the density of the normal-fluid
component.

The situation completely changes when we tune the gas parameter to
the strongly interacting regime (i.e., $n_{B}a_{B}^{3}=0.01$) \citep{Hu2010NJP}.
As shown in Fig. \ref{fig7_BECSoundVelocity}(b), the velocities of
first and second sound never become degenerate (cross) and thus the
hybridization found in Fig. \ref{fig7_BECSoundVelocity}(a) does not
occur. In this strongly interacting Bose superfluid, we also show
the approximate velocities of $c_{1}=v_{S}$ and $c_{2}=v/\sqrt{c_{P}/c_{V}}$,
given by the leading term of Eq. (\ref{eq:c1}) and Eq. (\ref{eq:c2}).
These approximate values give a reasonable first order approximation
to the full solutions of the first and second sound velocities. Overall,
we see that the qualitative behavior of sound velocities in the strongly
interacting Bose superfluid is close to that of a unitary Fermi gas
reported in Fig. \ref{fig4_SoundVelocity}.

\begin{figure}
\centering{}\includegraphics[width=1\columnwidth]{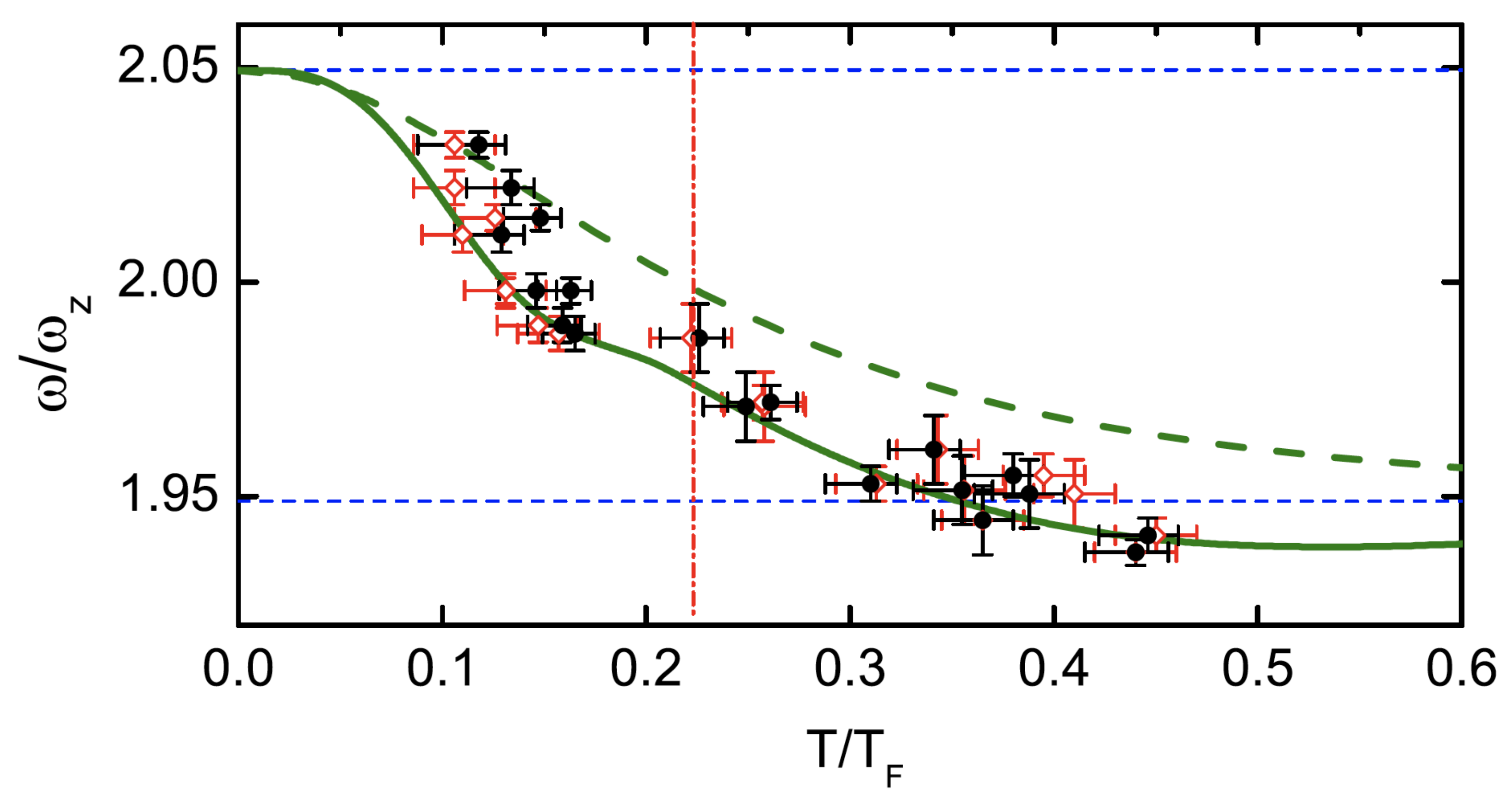}\caption{\label{fig8_LandauHD1D} Comparison between experimental (symbols)
and theoretical (green lines) first sound frequencies of the $k=2$
mode. The different symbols correspond to different ways of determining
temperature. In theoretical calculations, both the equations of state
of a unitary Fermi gas (green solid line) and of an ideal Fermi gas
(green dashed line) are used. The two thin blue horizontal dashed
lines mark the zero-temperature superfluid limit ($\omega/\omega_{z}=\sqrt{21/5}$)
and the classical hydrodynamic limit ($\omega/\omega_{z}=\sqrt{19/5}$),
respectively. The red dash-dotted vertical line indicates the critical
temperature in traps $T_{c}\simeq0.223(15)T_{F}$. Adapted from Ref.
\citep{Tey2013PRL}.}
\end{figure}

\section{Second sound in unitary Fermi gas: The first observation}

The analysis of hydrodynamic modes in an isotropic harmonic trap is
difficult to experimentally verify, since in experiments the trapping
potential is general axially symmetric. To overcome this problem,
Sandro Stringari and his colleagues suggested using a highly-elongated
cigar-like trap and derived reduced one-dimensional (1D) Landau's
two-fluid hydrodynamic equations \citep{Bertaina2010PRL,Hou2013PRA}.
Their brilliant idea on quasi-1D hydrodynamic modes eventually leads
to the first experimental observation of the second sound propagation
in 2013 \citep{Sidorenkov2013Nature}.

\subsection{Quasi-1D Landau's two-fluid hydrodynamic equations}

Let us consider that a unitary Fermi gas flows through a highly-elongated
trapping potential $V_{ext}(\mathbf{r})=m(\omega_{\perp}^{2}x^{2}+\omega_{\perp}^{2}y^{2}+\omega_{z}^{2}z^{2})/2$
with trapping frequency $\omega_{\perp}\gg\omega_{z}$. This case
may be compared with the flow of superfluid helium through a narrow
capillary tube \citep{Khalatnikov2000Book}. If the wavelength of
the helium flow is comparable to or greater than the diameter of the
tube, the normal-fluid component becomes stationary and sticks to
the hard wall of the tube due to its viscosity. The sound propagation
is then all carried by the superfluid component, forming the so-called
fourth sound \citep{Atkins1959PR}. In the current situation of a
\emph{soft} harmonic trap, the normal-fluid component will keep moving.
However, the viscosity may be sufficiently large to ensure that the
normal velocity $v_{n}(z)$ does not depend on the radial coordinates
($x$ or $y$), otherwise the sound propagation will involve excitations
costing large energy at the order of $\omega_{\perp}$ \citep{Bertaina2010PRL}.
Similarly, a large thermal conductivity may make the temperature uniformly
distributed in the radial direction \citep{Bertaina2010PRL}.

Due to the independence of the dynamic variables on the radial coordinates,
we may perform the integration over the radial degree of freedom in
all the equilibrium thermodynamic functions and deduce the quasi-1D
thermodynamics. For example, we can define the reduced 1D superfluid
density $n_{s1}(z)\equiv\int dxdyn_{s}(\mathbf{r})$, normal density
$n_{n1}(z)\equiv\int dxdyn_{n}(\mathbf{r})$, and entropy $s_{1}(z)\equiv\int dxdys(\mathbf{r})$.
By introducing the displacement fields $u_{s}(z)$ and $u_{n}(z)$,
we write down the variational action \citep{Hou2013PRA,Liu2014PRA},
\begin{eqnarray}
\mathcal{S}_{1D} & = & \frac{1}{2}\int dz\left[m\omega^{2}\left(n_{s1}u_{s}^{2}+n_{n1}u_{n}^{2}\right)-\left(\frac{\partial\mu}{\partial n_{1}}\right)_{s_{1}}\left(\delta n\right)^{2}\right.\nonumber \\
 &  & \left.-2\left(\frac{\partial T}{\partial n_{1}}\right)_{s_{1}}\delta n\delta s-\left(\frac{\partial T}{\partial s_{1}}\right)_{n}\left(\delta s\right)^{2}\right].\label{eq:usunAction1D}
\end{eqnarray}
where $n(z)=n_{s1}(z)+n_{n1}(z)$ is the reduced 1D total density,
and $\delta n(z)\equiv-\partial(n_{s1}u_{s}+n_{n1}u_{n})/\partial z$
and $\delta s(z)\equiv-\partial(s_{1}u{}_{n})/\partial z$ are the
density and entropy fluctuations, respectively. The effect of the
weak axial trapping potential $V_{ext}(z)=m\omega_{z}^{2}z^{2}/2$
now enters Eq. (\ref{eq:usunAction1D}) via the $z$-dependence of
the equilibrium thermodynamic variables, within the local density
approximation.

\begin{figure*}
\centering{}\includegraphics[width=0.9\textwidth]{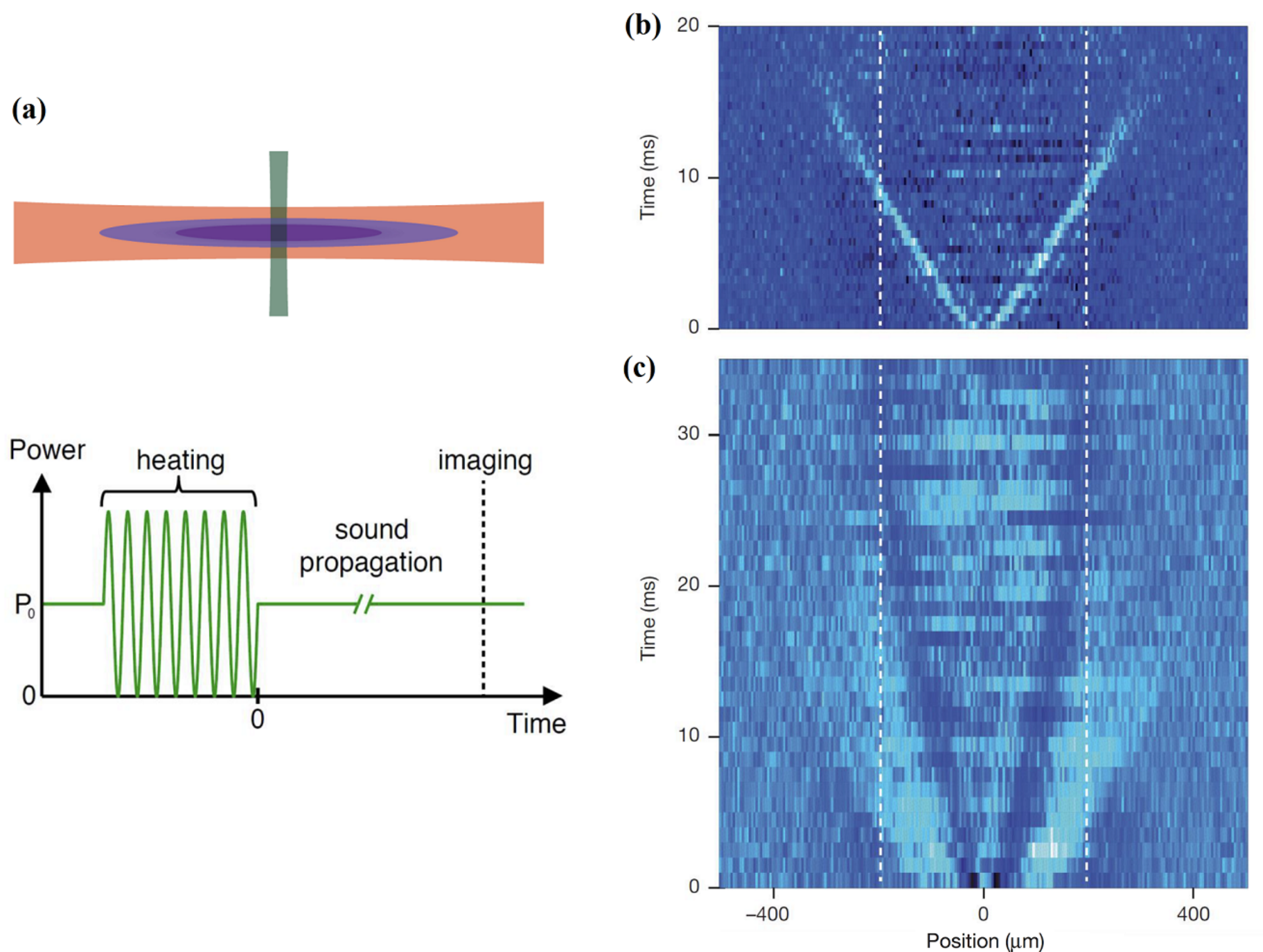}\caption{\label{fig9_InnsbruckSetup} (a) A second sound wave is excited by
a weak, power-modulated repulsive laser beam (green) that perpendicularly
intersects the cigar-like Fermi cloud (purple) with a total $N=3\times10^{5}$
$^{6}$Li atoms. The short power-modulation burst contains eight sinusoidal
oscillations (i.e., heating) in $4.5$ ms. The sinusoidal oscillations
are absent if a first sound wave is excited. (b) and (c) Normalized
differential axial density profiles, measured for variable delay times
after the excitation, show the propagation of first sound (local density
increase, bright) and second sound (local decrease, dark). The temperature
of the Fermi cloud is $T=0.135(10)T_{F}$, where $T_{F}=\hbar(3N\omega_{\perp}^{2}\omega_{z})^{1/3}$
is the Fermi temperature in traps. The vertical dashed lines indicate
the axial region where superfluid is expected to exist (i.e., $\left|z\right|<190$
$\mu$m). Adapted from Ref. \citep{Sidorenkov2013Nature}.}
\end{figure*}

To examine the reduced 1D two-fluid hydrodynamics, low-lying first
sound modes in elongated harmonic traps were measured by the Innsbruck
group \citep{Tey2013PRL}. First sound is the pure-phase mode with
$u_{s}(z)=u_{n}(z)=u^{(1)}(z)$. The minimization of the action $\mathcal{S}_{1D}$
with respect to $u^{(1)}(z)$ yields the following equation for the
first sound mode \citep{Tey2013PRL,Hou2013PRA}:
\begin{equation}
m\left(\omega^{2}-\omega_{z}^{2}\right)u^{(1)}=-\frac{7P_{1}}{5n_{1}}\frac{\partial^{2}u^{(1)}}{\partial z^{2}}+\frac{7}{5}m\omega_{z}^{2}z\frac{\partial u^{(1)}}{\partial z},\label{eq:FirstsoundTrap1D}
\end{equation}
where $P_{1}\equiv\int dxdyP(\mathbf{r})$ is the reduced 1D pressure.
It is easy to find the solution of Eq. (\ref{eq:FirstsoundTrap1D}),
by again using a polynomial \emph{Ansatz} \citep{Tey2013PRL,Hou2013PRA},
\begin{equation}
u^{(1)}(z)=a_{k}z^{k}+a_{k-2}z^{k-2}+\cdots
\end{equation}
with integer values of $k$. The lowest axial breathing mode has a
frequency $\omega_{k=1}=\sqrt{12/5}\omega_{z}$, which is temperature
independent due to the scaling invariance of the unitary Fermi gas
\citep{Ho2004PRL}. Experimentally, therefore it is useful to measure
the $k=2$ mode or $k=3$ mode, which shows the interaction effects
through the temperature dependence of the equation of state (EoS).
The theoretical prediction on the frequency of the $k=2$ first sound
mode as a function of temperature is presented in Fig. \ref{fig8_LandauHD1D}.
Indeed, by comparing the green solid line (with a unitary gas EoS
measured by the MIT group \citep{Ku2012Science}) and the green dashed
line (with an ideal Fermi gas EoS), we can see clearly the interaction
effect. The first sound mode frequency changes smoothly across the
superfluid phase transition (as indicated by the red dash-dotted vertical
line). The experimental data agree reasonably well with the theoretical
prediction based on the reduced 1D two-fluid hydrodynamic equations.
This provides a strong support for the applicability of the reduced
1D description.

We note that, the full minimization of the variational action with
respect to $u_{s}(z)$ and $u_{n}(z)$ has also been performed \citep{Liu2014PRA}.
The coupling between first sound and second sound only leads to a
very small correction to the $k=2$ mode frequency $\omega_{k=2}$.

\subsection{First observation of second sound propagation}

The Innsbruck group then considered the second sound propagation in
an elongated harmonic trap with trapping frequency $\omega_{\perp}=2\pi\times539(2)$
Hz and $\omega_{z}=2\pi\times22.46(7)$ Hz \citep{Sidorenkov2013Nature}.
For sound waves with excitation energy $\omega_{\perp}>\omega\gg\omega_{z}$,
the discreteness in energy due to the weak confining potential $V_{ext}(z)=m\omega_{z}^{2}z^{2}/2$
is not important. The sound waves propagate with a \emph{local} velocity
that slowly varies across the whole trap. By setting a plane wave
solution in the form of $e^{i(kz-\omega t)}$, it is straightforward
to derive the second sound velocity \citep{Hou2013PRA},
\begin{equation}
c_{2}=\sqrt{\frac{n_{s1}}{n_{n1}}\frac{T\bar{s}_{1}^{2}}{m\bar{c}_{P1}},}\label{eq:c21D}
\end{equation}
where the reduced 1D entropy per particle $\bar{s}_{1}$ and the reduced
1D specific heat per particle at constant pressure $\bar{c}_{P1}$
can be calculated by using the known EoS of a unitary Fermi gas, as
measured by the MIT group in 2012 \citep{Ku2012Science}. If the local
second sound velocity $c_{2}(z)$ is measured, from the above expression
one can then straightforward to determine the local reduced 1D superfluid
fraction $[n_{s1}/n_{1}](z)$ as a function of the local reduced temperature
$T/T_{F}^{1D}$, where $T_{F}^{1D}\equiv(15\pi/8)^{2/5}(\hbar\omega_{\perp})^{4/5}[\hbar^{2}n_{1}^{2}/(2m)]^{1/5}/k_{B}$
is a natural definition for Fermi temperature in highly-elongated
harmonic traps. Consequently, the superfluid fraction $n_{s}/n$ as
a function of the reduced temperature $T/T_{F}$ can be reconstructed,
thanks to the universal thermodynamic functions in the unitary limit
\citep{Ho2004PRL,Hu2007NaturePhysics}.

\begin{figure}
\centering{}\includegraphics[width=1\columnwidth]{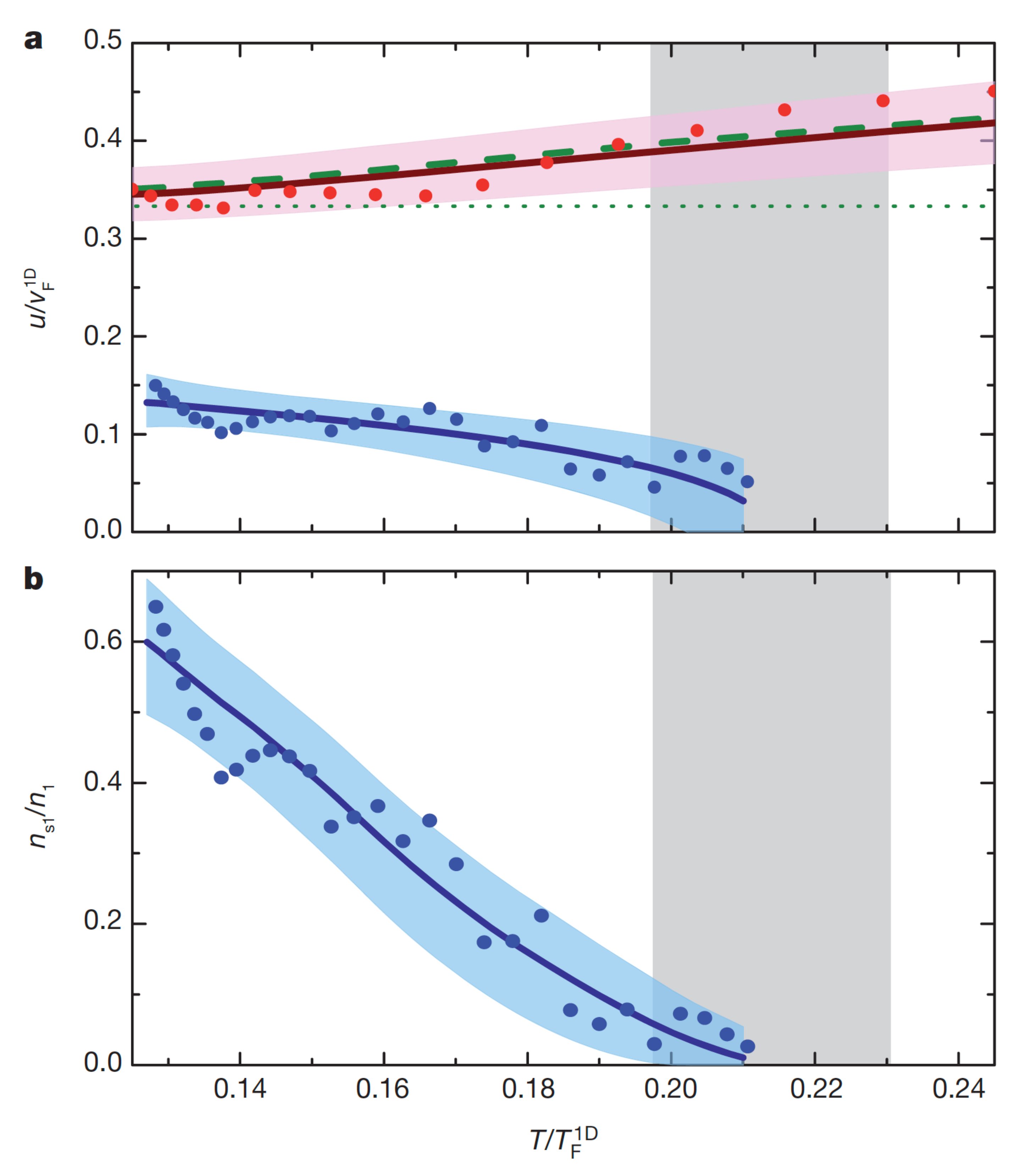}\caption{\label{fig10_InnsbruckSoundVelocity1D} (a) The measured first and
second sound velocities as a function of the reduced temperature $T/T_{F}^{1D}$.
The solid lines and symbols are experimental results from different
methods of taking the time derivative of the pulse position of sound
wave (i.e., through a polynomial fit or through analyzing subsets
of nine adjacent profiles). The green dashed line for first sound
velocity is the theoretical prediction $c_{1}=\sqrt{7P_{1}/(5mn_{1})}$.
(b) Temperature dependence of the reduced 1D superfluid fraction $n_{s1}/n_{1}$.
The grey shaded area indicates the uncertainty range of the superfluid
phase transition. From Ref. \citep{Sidorenkov2013Nature}.}
\end{figure}

To excite the second sound, the Innsbruck group used the same heating
strategy as in the superfluid helium. As shown in Fig. \ref{fig9_InnsbruckSetup}(a),
a weak power-modulated repulsive laser beam intersects the cigar-shape
Fermi cloud in the middle. A burst with eight sinusoidal oscillations
in a few milliseconds then excites a second sound wave, whose propagation
with time is presented in Fig. \ref{fig9_InnsbruckSetup}(c). Two
local density dips are clearly visible within the superfluid region
(as enclosed by the two vertical dashed lines). The dips move very
slowly when they approach the superfluid boundary and finally disappear.
The fact that this sound wave cannot propagate in the non-superfluid
region is a fingerprint characteristic of the second sound.

In contrast, a first sound wave can be excited without the heating
sinusoidal oscillations, as shown in Fig. \ref{fig9_InnsbruckSetup}(b).
The first sound manifests itself as two local density peaks. They
move much faster than the second sound dips in Fig. \ref{fig9_InnsbruckSetup}(c).
Obviously, they can penetrate through the superfluid boundary and
propagate freely into the non-superfluid region, as anticipated.

The first and second velocities can be easily extracted from the time
evolution of the peak position (for first sound) and the dip position
(for second sound). For example, by fitting the pulse (peak or dip)
position as function of time with a third-order polynomial, the sound
velocity can be straightforwardly determined by taking the time-derivative
of the fitting curve. The results are shown in Fig. \ref{fig10_InnsbruckSoundVelocity1D}(a)
by solid lines. Alternatively, the time-derivative can be taken by
analyzing subsets of nine adjacent profiles, which leads to the data
points (symbols) in the figure. For the first sound velocity, the
experimental data agree well with the theoretical prediction $c_{1}=\sqrt{7P_{1}/(5mn_{1})}$,
which can be easily derived by using Eq. (\ref{eq:FirstsoundTrap1D})
in the absence of the weak confining potential (i.e., $\omega_{z}=0$). 

From the second sound velocity, it is straightforward to determine
the reduced 1D superfluid fraction $n_{s1}/n_{1}$, by using Eq. (\ref{eq:c21D}).
This unknown quantity is presented in Fig. \ref{fig10_InnsbruckSoundVelocity1D}(b).
It decreases monotonically with increasing temperature and vanishes
in the uncertainty range of the superfluid phase transition.

\begin{figure}
\centering{}\includegraphics[width=1\columnwidth]{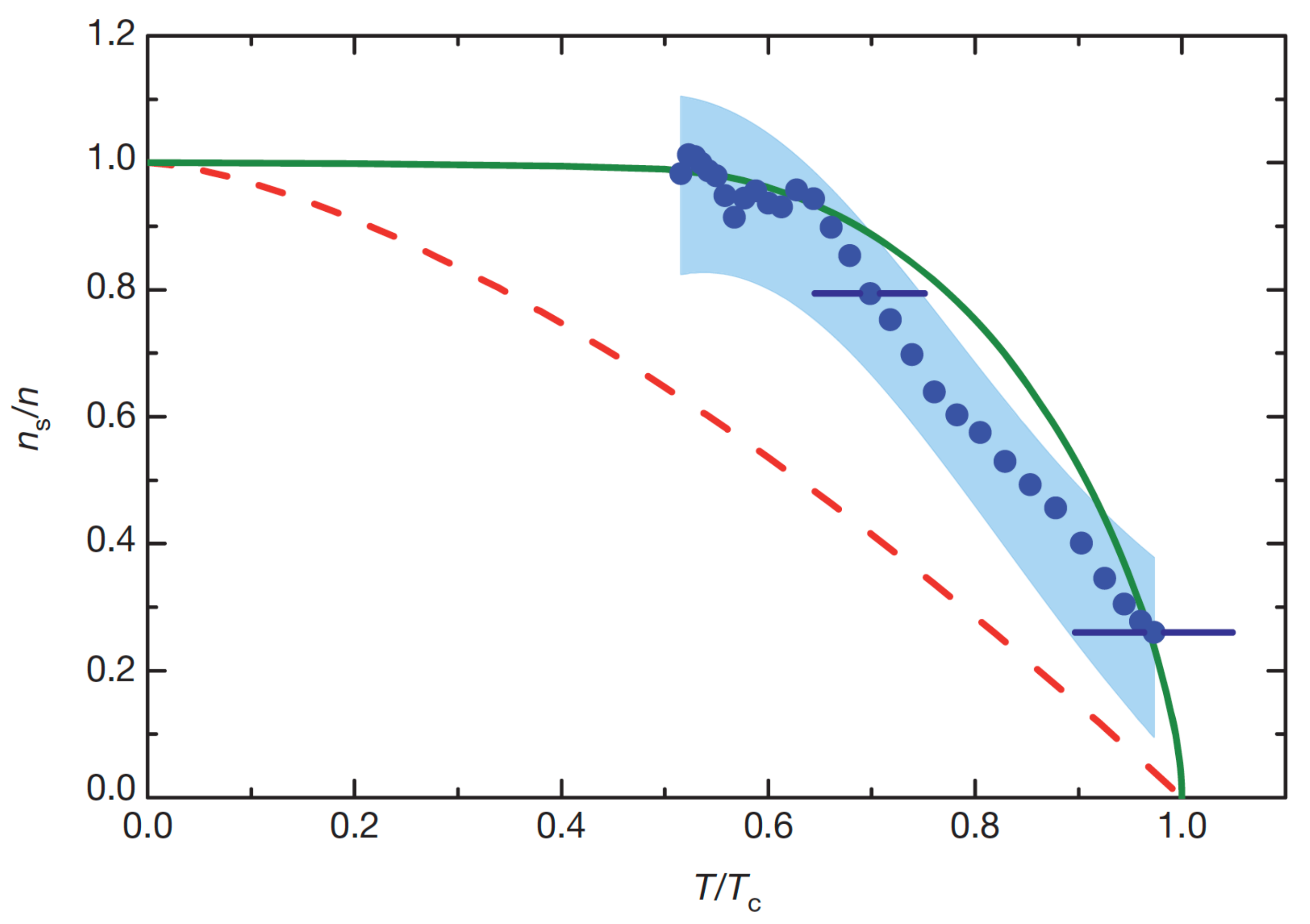}\caption{\label{fig11_nsInnsbruck} The reconstructed superfluid fraction of
a uniform unitary Fermi gas (symbols) as a function of $T/T_{c}$,
with uncertainty range indicated by shaded region. For comparison,
the superfluid fraction of superfluid helium (green line) and the
condensate fraction of an ideal Bose gas $1-(T/T_{c})^{3/2}$ (red
dashed line) are also shown. From Ref. \citep{Sidorenkov2013Nature}.}
\end{figure}

The reconstruction of the most important uniform superfluid fraction
$n_{s}/n$ from $n_{s1}/n_{1}$ is also straightforward \citep{Sidorenkov2013Nature}.
The result is shown in Fig. \ref{fig11_nsInnsbruck}, with the uncertainty
indicated by the blue shaded area. It is interesting to see that the
measured superfluid fraction of unitary Fermi gas follows closely
that of superfluid helium (green solid line). However, the uncertainty
of the data points seems to be too large to make a convincing conclusion.
Nevertheless, the superfluid fractions of both unitary Fermi gas and
superfluid helium are much larger than the condensate fraction of
an ideal Bose gas (red dashed line).

\begin{figure}
\centering{}\includegraphics[width=1\columnwidth]{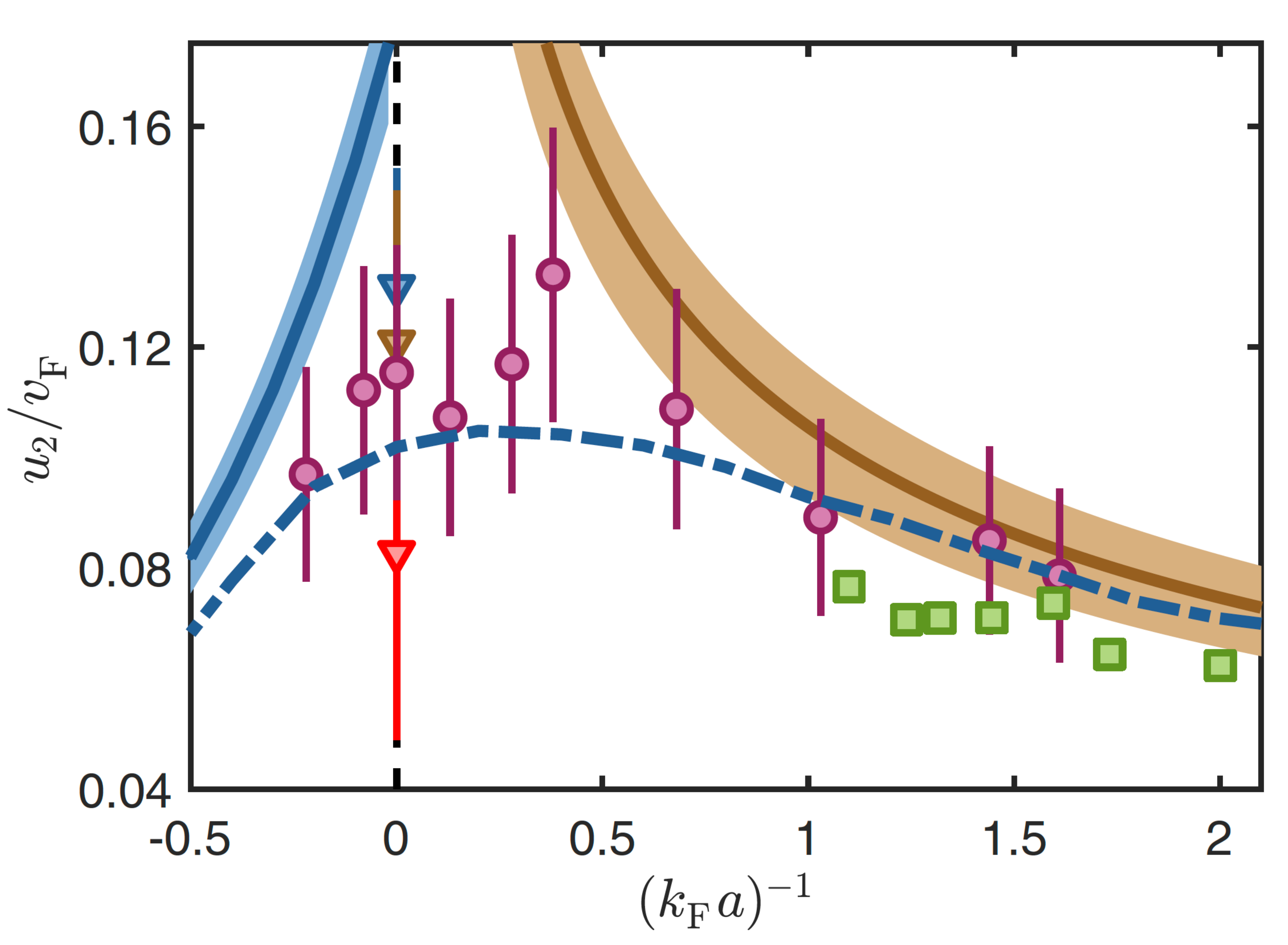}\caption{\label{fig12_c2Ulm} The purple circles are the measured data from
the Ulm experiment for temperatures in the range $T/T_{c}=0.66-0.84$.
The brown and blue solid line show analytic predictions for the BEC
and BCS regime at $T=0.75T_{c}$, respectively. The blue dash-dotted
line shows an earlier theoretical prediction of second sound at the
crossover at $T/T_{c}=0.75$ \citep{Heiselberg2006PRA}. The green
squares are results of numerical $c$-field simulations. For comparison,
the second sound velocities of a unitary Fermi gas measured by the
Innsbruck group \citep{Sidorenkov2013Nature} at the temperatures
$T/T_{c}=0.65$ (blue triangle), $T/T_{c}=0.75$ (brown triangle),
and $T/T_{c}=0.85$ (red triangle) are also shown. From Ref. \citep{Hoffmann2021NatureCommunications}.}
\end{figure}

\subsection{Second sound propagation at the BEC-BCS crossover}

The landmark experiment by the Innsbruck group focused on the unitary
limit. Recently, the same-type measurement has been carried out at
the whole BEC-BCS crossover by D. K. Hoffmann and his colleagues at
Ulm University \citep{Hoffmann2021NatureCommunications}. Their result
of the second sound velocity as a function of the interaction parameter
$1/(k_{F}a)$ is presented in Fig. \ref{fig12_c2Ulm}. The temperature
varies in the range $T/T_{c}=0.66-0.84$. In this temperature range,
the second sound velocity shows a pronounced maximum slightly above
the unitary limit $1/(k_{F}a)\sim0.4$. Overall, the measured second
sound velocity agrees qualitatively well with an earlier theoretical
calculation (blue dash-dotted line) based on the mean-field superfluid
density and thermodynamic functions \citep{Heiselberg2006PRA}. A
refined theoretical analysis beyond the mean-field treatment would
be useful to better understand the experimental data at the BEC-BCS
crossover.

\section{Second sound in unitary Fermi gas: The measurement of sound attenuation}

It is a milestone to observe the second sound of unitary Fermi gas
in highly-elongated harmonic traps \citep{Sidorenkov2013Nature}.
However, the trapping potential seems to be very unfriendly for the
purpose of performing precise measurements. On the one hand, the variation
of local density along the weakly-confined direction brings a relatively
large uncertainty on the measured 1D sound velocities and hence on
the extracted 1D superfluid fraction, as we already see from Fig.
\ref{fig10_InnsbruckSoundVelocity1D}. On the other hand, the reduced
1D Landau's two-fluid hydrodynamic equations are dissipationless due
to the dimensional reduction \citep{Bertaina2010PRL,Hou2013PRA}.
As a result, there is no theoretical guidance on how to extract the
important second sound attenuation and the related transport coefficients
such as shear viscosity $\eta$ and thermal conductivity $\kappa$,
if the damping of 1D sound propagation is measured.

This serious technical difficulty could be overcome, owing to the
experimental realization of a box-like trapping potential and consequently
a homogeneous interacting Fermi or Bose gas \citep{Gaunt2013PRL,Mukherjee2017PRL}.
Indeed, the first sound attenuation of a homogeneous unitary Fermi
gas was recently measured by P. B. Patel and his colleagues at Massachusetts
Institute of Technology (MIT) \citep{Patel2020Science}. The measurement
of second sound and its attenuation has also been attempted at MIT,
by setting up a local cold-atom thermometry to measure the temperature
wave. However, this is difficult. It turns out that a direct and more
efficient way is to measure again the density signal of second sound
thanks to the appreciable LP ratio of unitary Fermi gas. A high-resolution
Bragg spectroscopy was set up for this purpose by the USTC group \citep{Li2022Science}.
The second sound attenuation was then successfully measured \citep{Li2022Science}.

\subsection{Hydrodynamic density response function}

Before we discuss the USTC experiment on the second sound attenuation,
it is useful to briefly describe the density response function in
Landau's two-fluid hydrodynamic theory. Its expression was first derived
by Hohenberg and Martin in their seminal work \citep{Hohenberg1965AoP}
and takes the following form for an isotropic superfluid (such that
$\chi(\mathbf{k},\omega)=\chi(k,\omega)$),
\begin{equation}
\chi=\frac{\left(nk^{2}/m\right)\left(\omega^{2}-v^{2}k^{2}+iD_{s}k^{2}\omega\right)}{\left(\omega^{2}-c_{1}^{2}k^{2}+iD_{1}k^{2}\omega\right)\left(\omega^{2}-c_{2}^{2}k^{2}+iD_{2}k^{2}\omega\right)},\label{eq:DissipativeDensityReponse}
\end{equation}
where the damping rate or sound attenuation of the response is characterized
by three diffusion coefficients $D_{1}$, $D_{2}$ and $D_{s}$, as
determined by solving the coupled equations,
\begin{eqnarray}
D_{1}+D_{2} & = & \frac{4\eta}{3\rho}\frac{n}{n_{n}}+\frac{\kappa}{\rho c_{V}},\label{eq: DC1}\\
\frac{c_{1}^{2}D_{2}+c_{2}^{2}D_{1}}{v_{S}^{2}} & = & \frac{4\eta v^{2}}{3\rho v_{S}^{2}}\left[1-\frac{2\left(\frac{\partial P}{\partial T}\right)_{n}}{sn}+\frac{n_{s}v_{S}^{2}}{n_{n}v^{2}}\right]+\frac{\kappa}{\rho c_{P}},\label{eq: DC2}\\
D_{s} & = & \frac{4\eta}{3\rho}\frac{n_{s}}{n_{n}}+\frac{\kappa}{\rho c_{V}}.\label{eq: DC3}
\end{eqnarray}
Here $\rho=mn=m(n_{s}+n_{n})$ is the total mass density, and we do
not include the various second viscosities $\zeta_{i}$ ($i=1$,$2$,$3$,$4$),
since for a unitary Fermi gas only $\zeta_{3}$ can be nonzero but
its value could be insignificant to have sizable contribution. In
the absence of the diffusivities $D_{1}$, $D_{2}$ and $D_{s}$,
the hydrodynamic density response Eq. (\ref{eq:DissipativeDensityReponse})
reduces to the familiar dissipationless form in Eq. (\ref{eq:dissipationlessDensityResponse}).
There are two sources for dissipation that is responsible for the
sound attenuation. One is the viscous damping stemming from the diffusion
of momentum, characterized by the shear viscosity $\eta$ (and various
second viscosities that we neglect). Another is the thermal damping
caused by diffusion of heat, given by the thermal conductivity $\kappa$.
The relative contribution of these two sources can be quantified by
a dimensionless Prandtl number,
\begin{equation}
\textrm{Pr}\equiv\frac{\eta c_{P}}{\kappa}.
\end{equation}

It is interesting to note that, over the last two decades, the measurements
of both shear viscosity $\eta$ and thermal conductivity $\kappa$
of strongly interacting unitary Fermi gas have received increasing
attention from a diverse fields of physics. For shear viscosity $\eta$,
it is anticipated that the unitary Fermi gas may behave like a \emph{perfect}
fluid, with viscosity-to-entropy ratio $\eta/s$ close to a universal
lower-bound $1/(4\pi k_{B})$ conjectured by Kovtun, Son, and Starinets
(KSS) \citep{Kovtun2005PRL}. The measurement of thermal conductivity
$\kappa$ is equally important. From the AdS/CFT correspondence, a
holographic conformal nonrelativistic fluid has a Prandtl number $\textrm{Pr}=1$
\citep{Ross2009JHEP}. Therefore, it would be interesting to check
whether a unitary Fermi gas can be treated as a holographic dual near
the superfluid phase transition, where the conformal invariance might
be satisfied. We also note that, the shear viscosity of unitary Fermi
gas has been measured at North Carolina State University (NCSU) from
the anisotropic expansion of the Fermi cloud for a wide temperature
window \citep{Joseph2015PRL}, ranging from $\sim T_{F}$ down to
far below the transition temperature. Less is known about the thermal
conductivity, except the high-temperature classical regime \citep{Braby2010PRA},
where $\kappa_{\textrm{cl}}=(n\hbar k_{B}/m)(675\sqrt{2}\pi^{3/2}/512)(T/T_{F})^{3/2}$.

As in the dissipationless case, it is intuitive to separate the first
and second sound contributions to the density response function Eq.
(\ref{eq:DissipativeDensityReponse}). By ignoring the high-order
terms of the small parameter $\epsilon_{\textrm{LP}}(c_{2}/c_{1})^{2}$,
we find that \citep{Li2022Science},
\begin{equation}
\chi\left(k,\omega\right)=\chi^{(1)}\left(k,\omega\right)+\bar{\chi}^{(2)}\left(k,\omega\right)+\cdots,
\end{equation}
where
\begin{eqnarray}
\chi^{(1)}\left(k,\omega\right) & = & \frac{nk^{2}}{m}\frac{\left(c_{1}^{2}-v^{2}\right)/\left(c_{1}^{2}-c_{2}^{2}\right)}{\omega^{2}-c_{1}^{2}k^{2}+iD_{1}k^{2}\omega},\label{eq:kappa1}\\
\bar{\chi}^{(2)}\left(k,\omega\right) & = & \epsilon_{\textrm{LP}}\frac{nk^{2}}{mc_{1}^{2}}\frac{c_{2}^{2}-iD_{T}\omega}{\omega^{2}-c_{2}^{2}k^{2}+iD_{2}k^{2}\omega},\label{eq:kappa2}
\end{eqnarray}
with the thermal diffusivity $D_{T}\equiv\kappa/(\rho c_{P})$. It
is easy to see that, above the superfluid transition ($c_{1}=v_{S}$
and $D_{2}=D_{T}$), the second sound turns into a thermally diffusive
mode, with a contribution \citep{Hu2018PRA}
\begin{equation}
\bar{\chi}^{(2)}\left(k,\omega\right)=-\epsilon_{\textrm{LP}}\frac{n}{mv_{S}^{2}}\frac{iD_{T}k^{2}}{\omega+iD_{T}k^{2}},\label{eq:kappa2p}
\end{equation}
which peaks around the zero frequency $\omega=0$.

\begin{figure*}
\centering{}\includegraphics[width=0.9\textwidth]{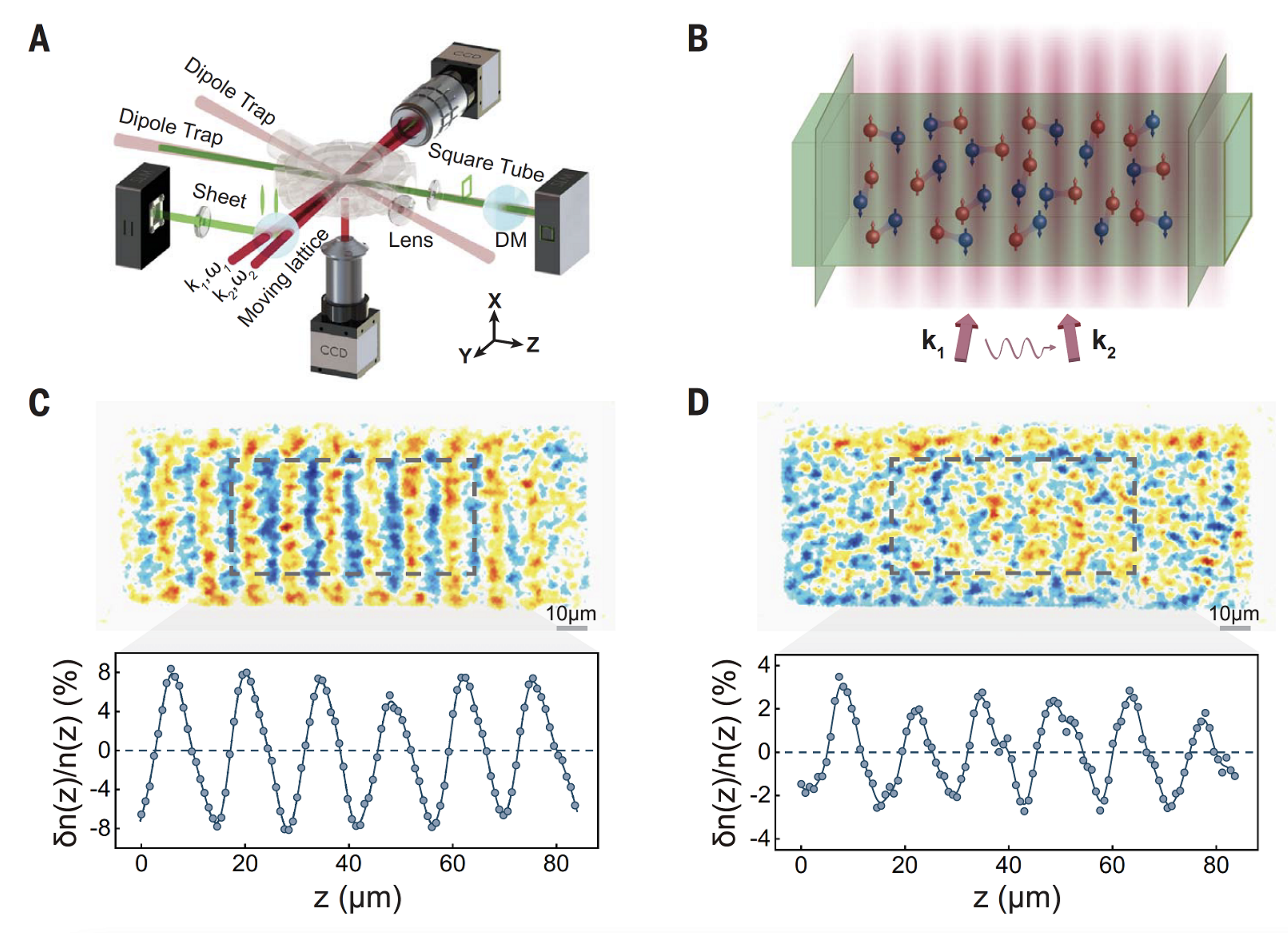}\caption{\label{fig13_USTCsetup} (a) The experimental setup: a box trap consists
of a square-tube and two sheets of laser beams. (b) A pair of 741
nm laser beams with wavenumbers ($k_{1}=k_{2}=k_{L}$) and frequencies
($\omega_{1},\omega_{2}$) intersect on the homogeneous unitary Fermi
gas with a small angle, producing a one-dimensional moving-lattice
potential. (c) and (d) First and second sound waves. The upper panels
show two typical single-shot images of unitary superfluid at $T=0.84T_{c}$,
taken at the frequencies $\omega=\omega_{1}-\omega_{2}=2.1$ kHz and
$0.3$ kHz, respectively. To clearly show the density waves along
the longitudinal $z$-axis, we have taken the difference with respect
to a reference image at $\omega=1$ MHz. Further integration of the
difference along the two transverse directions in the region of interest,
as marked by the dashed box, reveals a normalized density fluctuation
$\delta n/n$ propagating along the $z$-axis, as shown in the low
panels. From Ref. \citep{Li2022Science}.}
\end{figure*}

\subsection{Second sound attenuation measurement}

In the USTC experiment \citep{Li2022Science}, a novel high-resolution
Bragg spectroscopy has been developed to directly probe the density
signals from both first sound and second sound. As shown in Fig \ref{fig13_USTCsetup}(a),
a high-density population-balanced Fermi mixture at the Feshbach resonance
is realized in a box trap with about $10^{7}$ $^{6}$Li atoms, much
denser than earlier experiments where the number of atoms is typically
in the range of $10^{4}-10^{6}$. This state-of-the-art preparation
of Fermi superfluid typically yields a high density $n\simeq1.56\times10^{13}~\text{cm}^{-3}$,
a large Fermi wavenumber $k_{F}=(3\pi^{2}n)^{1/3}\simeq2\pi\times1.23~\mu$m$^{-1}$,
and also a large Fermi energy $E_{F}\simeq2\pi\hbar\times50$ kHz.
By further applying a pair of coherent 741 nm laser beams with frequency
difference $\omega$ (see Fig. \ref{fig13_USTCsetup}(b)), a moving-lattice
potential is formed to create density fluctuations. The angle between
two lattice lasers can be finely tuned to give rise to a very small
Bragg momentum $k=2\pi\times0.071~\mu$m$^{-1}\simeq0.058~k_{F}$.
Therefore, if we estimate the correlation length $\xi\sim k_{F}^{-1}|t|^{-\nu}$
as in superfluid helium and optimistically take the criterion $k\xi<1$
for the hydrodynamic regime, the hydrodynamic description for the
density response function Eq. (\ref{eq:DissipativeDensityReponse})
would be applicable over a wide temperature range unless very close
to the superfluid transition, i.e., $|t|<0.014$ or $|T-T_{c}|<0.002~T_{F}$.
After applying the moving lattice potential with strength $V_{0}$
for 3 milliseconds, a steady-state density fluctuation can be measured,
as reported in Fig. \ref{fig13_USTCsetup}(c) for first sound and
Fig. \ref{fig13_USTCsetup}(d) for second sound at $T=0.84T_{c}$.
The density fluctuation should take the form, $\delta n(z,t)=|\chi(k,\omega)|V_{0}\sin[kz-\omega t+\varphi(k,\omega)]$,
from which the modulus of the density response function $|\chi(k,\omega)|$
is determined.

There are two key technical advantages of the developed high-resolution
Bragg spectroscopy at USTC. First, the modulus $|\chi(k,\omega)|$
is directly obtained from the integration of $\left|\delta n/n\right|$
as a function of $\omega$, contrasted with $\text{Im}[\chi(k,\omega)]$
from the out-of-phase density response in other experiments \citep{Christodoulou2021Nature,Patel2020Science}.
This avoids potential errors due to the phase synchronization. On
the other hand, more importantly, the steady-state density response
removes a large instrumental spectral broadening $\sim0.1E_{F}$ due
to a finite Bragg pulse duration in the conventional Bragg spectroscopy
\citep{Kuhnle2010PRL,Zou2010PRA} (as well as in the inelastic neutron
scattering experiments of superfluid helium \citep{Griffin1993Book}).
Both advantages, together with the large Fermi energy, makes it possible
to directly probe the second sound.

\begin{figure}
\centering{}\includegraphics[width=1\columnwidth]{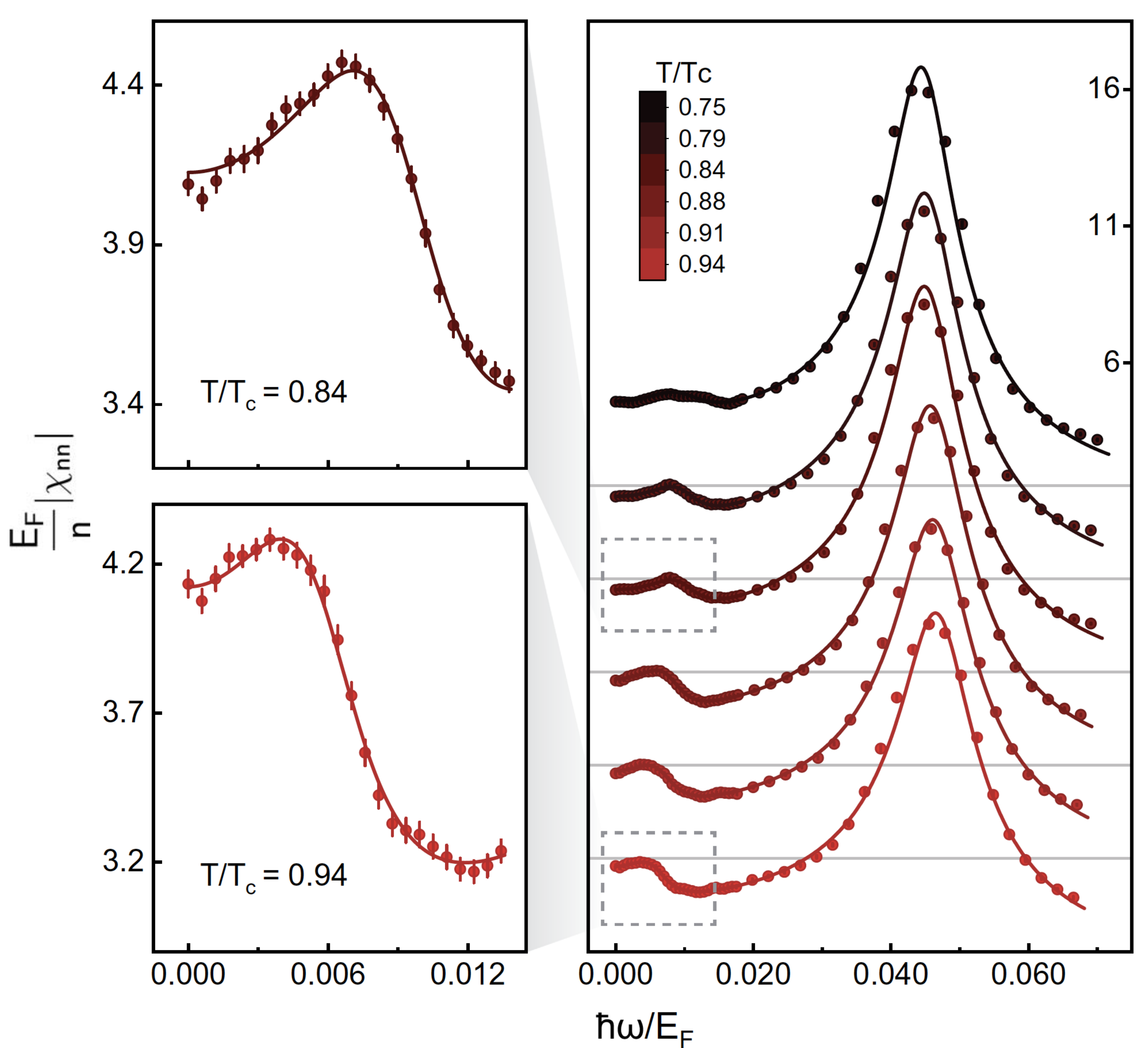}\caption{\label{fig14_USTCDensityResponse} The spectra from $0.75T_{c}$ to
$0.94T_{c}$ (top to bottom) are shown on the right panel. The two
small subplots on the left give a zoomed view of the low-lying second-sound
response at $0.84T_{c}$ and $0.94T_{c}$, respectively. The solid
lines are the fitting curves, in the form of Eq. (\ref{eq:DissipativeDensityReponse}).
Adapted from Ref. \citep{Li2022Science}.}
\end{figure}

The density response spectra $\left|\chi(k,\omega)\right|$ over a
wide range of temperature $T\subseteq[0.42,1.04]T_{c}$ have been
systematically measured in the experiment. The representative spectra
from 0.75$T_{c}$ to 0.94$T_{c}$ are reported in Fig. \ref{fig14_USTCDensityResponse}.
A second sound peak can be clearly identified at low-frequency $\omega<0.01E_{F}$,
in addition to a high-frequency first-sound peak at $\omega\sim0.045E_{F}$.
The simultaneous appearance of both first sound and second sound signals
is a unambiguous proof of Landau's two-fluid description of the density
response in the long-wavelength and low-energy hydrodynamic regime.
Indeed, the measured density response can be well fitted by using
Eq. (\ref{fig14_USTCDensityResponse}), with various sound velocities
($c_{1}$, $c_{2}$ and $v$) and sound diffusivities ($D_{1}$, $D_{2}$
and $D_{s}$) as six fitting parameters.

The second sound signal is particularly evident at $T\sim0.8-0.9T_{c}$,
as highlighted in the left panel of Fig. ~\ref{fig14_USTCDensityResponse}.
It is worth noting that, the peak does not show a symmetric Lorentzian
line-shape as we anticipate for a propagating sound, since it is the
modulus $\left|\chi(k,\omega)\right|$ being plotted, instead of $\text{Im}[\chi(k,\omega)]$.
At zero frequency, the density response function is given by the compressibility
sum-rule \citep{Griffin1993Book,Hu2010NJP}, 
\begin{equation}
\chi\left(k,\omega=0\right)=-\frac{n}{mv_{T}^{2}},
\end{equation}
and therefore is nonzero. To further characterize the second sound
contribution in the density response, $\text{Im}[\chi(k,\omega)]$
can be reconstructed by using the six fitting parameters. A well-defined
propagating second-sound with Lorentzian shape is found, for temperature
up to a threshold $0.98T_{c}$ (not shown in the figure). This threshold
temperature seems to be consistent with the estimation for hydrodynamics
based on the criterion $k\xi<1$, which yields a characteristic temperature
$T\sim0.986T_{c}$ as discussed earlier.

\begin{figure}
\centering{}\includegraphics[width=1\columnwidth]{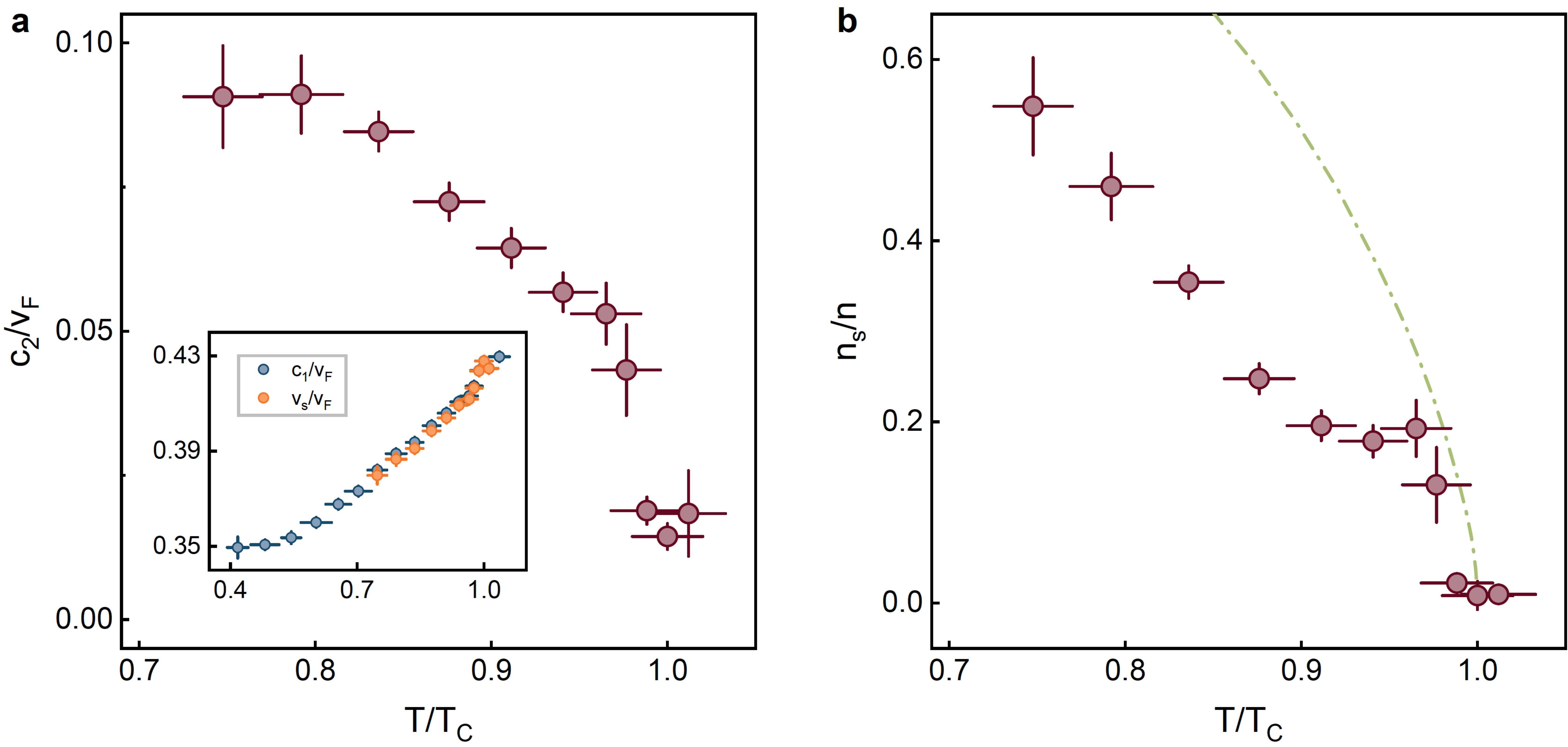}\caption{\label{fig15_nsUSTC} (a) Temperature dependence of the normalized
first-sound speed $c_{1}/v_{F}$ (blue dots in the inset) and second-sound
speed $c_{2}/v_{F}$ (red dots), obtained from the curve fitting of
density response spectra Fig.~\ref{fig14_USTCDensityResponse}. Here,
$v_{F}=\hbar k_{F}/m$ is the Fermi velocity. The inset also shows
the adiabatic sound speed $v_{S}/v_{F}$ (orange dots), calculated
by using $v_{S}=\sqrt{c_{1}^{2}+c_{2}^{2}-v^{2}}$. (b) Temperature
dependence of the superfluid fraction $n_{s}/n$, compared with that
of superfluid helium (green dash-dotted line). Adapted from Ref. \citep{Li2022Science}.}
\end{figure}

From the curve fitting, the high-quality density response spectra
give rise to fairly accurate results on the sound velocities, as presented
in Fig. \ref{fig15_nsUSTC} as a function of the reduced temperature
$T/T_{c}$. The accuracy of the first-sound speed $c_{1}$ and adiabatic
sound speed $v_{S}$ is impressive, with a typical relative error
of just $\sim1\%$. This provides a new route to measure the universal
equation of state of the unitary superfluid. For example, from the
saturated first-sound speed $c_{1}/v_{F}=0.350(4)$ at the lowest
temperature $0.42T_{c}$, the Bertsch parameter $\xi_{0}=0.367(9)$
could be deduced by using the standard thermodynamic relation $c_{1}/v_{F}=\sqrt{\xi_{0}/3}$.
This value agrees excellently well with the state-of-the-art quantum
Monte Carlo result \citep{Jensen2020PRL}, $\xi_{0}=0.367(7)$. On
the other hand, the obtained second sound velocity has a much smaller
uncertainty than the previous measurement at Innsbruck \citep{Sidorenkov2013Nature}.
The second sound speed depends sensitively on the temperature: $c_{2}/v_{F}$
decreases rapidly down to $0.98~T_{c}$ with increasing temperature
and then suddenly becomes negligible. The sudden decrease could be
related to the breakdown of the hydrodynamic description near the
phase transition.

By applying the well-known relation $v^{2}=(Ts^{2}n_{s})/(mC_{V}n_{n})$
or by using Eq. (\ref{eq:c2Approximate}), the superfluid fraction
$n_{s}/n$ can be directly determined, as shown in Fig.~\ref{fig15_nsUSTC}(b).
In the vicinity of the phase transition, the superfluid fraction is
close to that of superfluid helium, as found earlier at Innsbruck
\citep{Sidorenkov2013Nature}. By decreasing temperature to $0.97T_{c}$,
the unitary Fermi superfluid rapidly acquires about $20\%$ superfluid
component. As temperature further decreases, however, the data start
to significantly depart from the case of superfluid helium, in sharp
contrast to the earlier observation at Innsbruck.

\begin{figure}
\centering{}\includegraphics[width=1\columnwidth]{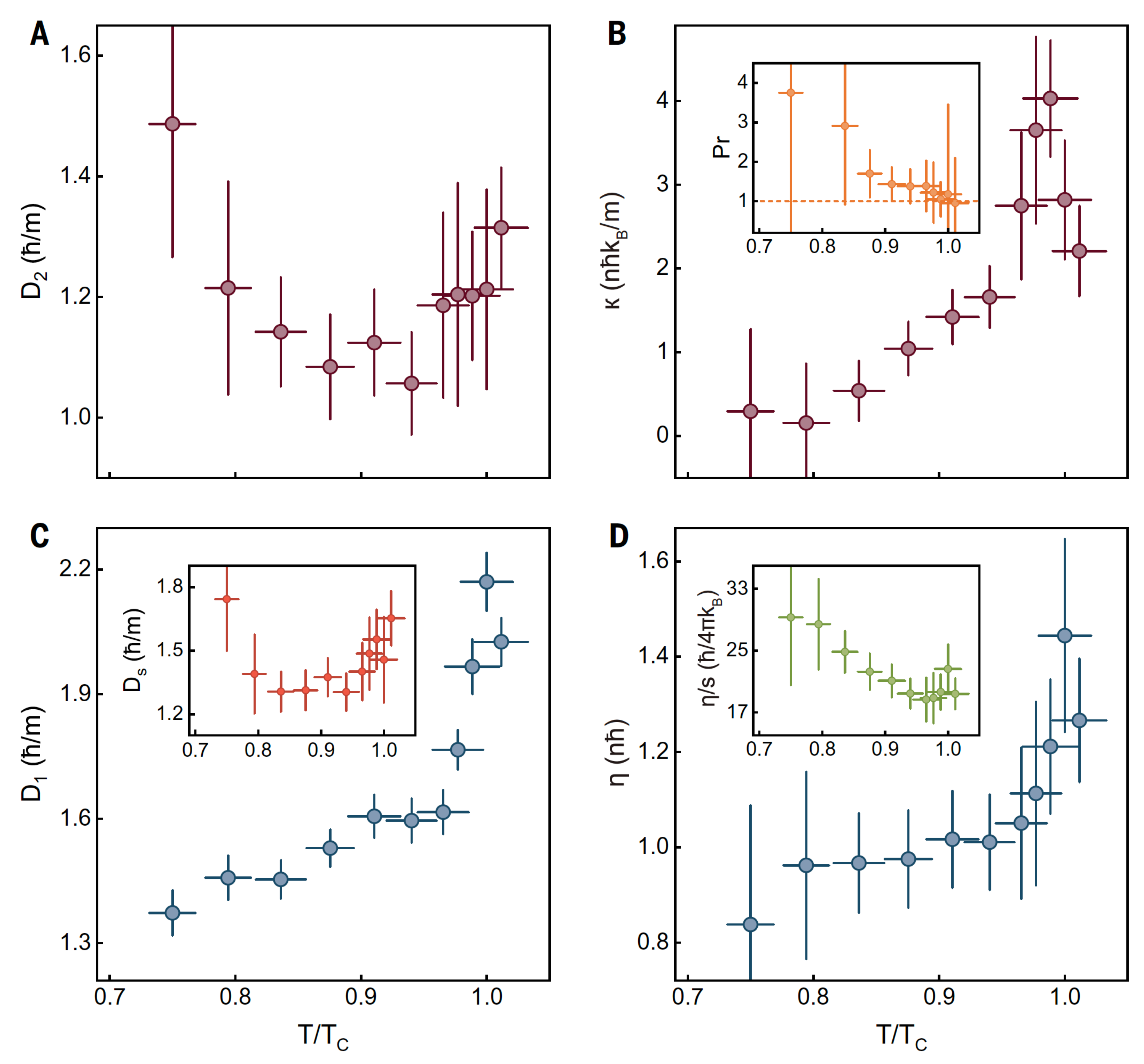}\caption{\label{fig16_USTCTransportCoefficients} (a) The second-sound diffusivity
$D_{2}$. (b) The thermal conductivity $\kappa$. The inset shows
the Prandtl number, with the line marking $\text{Pr}=1$. (c) The
first-sound diffusivity $D_{1}$, together with $D_{s}$ in the inset.
(d) The shear viscosity $\eta$. The inset reports the viscosity-to-entropy
ratio, in units of $\hbar/(4\pi k_{B})$. From Ref. \citep{Li2022Science}.}
\end{figure}

The key results of the USTC experiment - the first and second sound
attenuation - are presented in Fig.~\ref{fig16_USTCTransportCoefficients},
along with the extracted transport coefficients by using the relations
$\kappa\simeq[(D_{1}+D_{2})mn-4\eta n/(3n_{n})]C_{V}$ and $\eta=3nm(D_{1}+D_{2}-D_{s})/4$.
The temperature dependence of $D_{1}$ and $D_{2}$ below $T_{c}$
might be understood from the relations $D_{1}\sim\eta/(nm)$ and $D_{2}\sim\eta n_{s}/(nmn_{n})$
valid at low temperature \citep{Li2022Science}. As temperature decreases,
the saturated $D_{1}$ is consistent with a nearly constant shear
viscosity and the rapid increase of $D_{2}$ is due to the loss of
normal fluid component, i.e., $n_{s}/n_{n}\rightarrow\infty$ as $T\rightarrow0$.
Very close to the superfluid transition (i.e., at $T>0.95~T_{c}$),
all the sound diffusivities show a sudden rise, with a variation $\delta D\sim0.3\hbar/m$. 

In comparison with Fig. \ref{fig2_Helium2ndSoundAttenuation}, it
is remarkable that the second-sound diffusivity $D_{2}$ of a unitary
Fermi gas has a very similar temperature dependence as that of superfluid
helium. The sudden rise in $D_{2}$ near the superfluid transition
then could be understood as a critical divergence, as we discussed
earlier in superfluid helium section. Moreover, $D_{2}$ clearly shows
a minimum at $T\sim0.9T_{c}$. Although the minimum value $(D_{2})_{\min}\simeq1.1\hbar/m$
slightly smaller than the minimum $D_{2}$ of superfluid helium given
in Eq. (\ref{eq:D2minHeII}), the important observation is that, there
is a universal limit $D\sim\hbar/m$ for both strongly interacting
quantum fluids, due to the absence of well-defined quasi-particles. 

\subsection{Transport coefficients and quantum criticality}

As in superfluid helium, the critical divergence in the second-sound
diffusivity $D_{2}$ near the superfluid transition should be attributed
to the thermal conductivity $\kappa$, which diverges like $\left|T-T_{c}\right|^{-\nu/2}\simeq\left|T-T_{c}\right|^{-1/3}$
both below and above $T_{c}$, according to the dynamic critical scaling
theory of the superfluid transition \citep{Ferrell1967PRL,Hohenberg1977RMP}.
Indeed, as shown in Fig. \ref{fig16_USTCTransportCoefficients}(b),
there is a clear weak divergence on both sides of the superfluid transition,
leading a pronounced peak around $T_{c}$ with a significant height
$\delta\kappa\sim3n\hbar k_{B}/m$. The Prandtl number $\text{Pr}\equiv\eta c_{P}/\kappa$
is also reported in the inset. This number is about unity near the
superfluid transition, suggesting that there the viscous damping and
thermal damping are equally important to the sound attenuation. A
similar Prandtl number has been found using Brillouin scattering in
superfluid helium \citep{Tarvin1977PRB}. The near unity Prandtl number
also implies that we may treat the unitary Fermi gas as a holographic
conformal nonrelativistic fluid \citep{Ross2009JHEP}.

Finally, the shear viscosity is presented in Fig. \ref{fig16_USTCTransportCoefficients}(d).
It exhibits a weak temperature dependence below 0.95~$T_{c}$ and
becomes nearly constant, i.e., the quantum limit $\eta\sim n\hbar$.
In the vicinity of the superfluid transition, a smooth but significant
increase is observed. The trap-averaged shear viscosity of a unitary
Fermi gas in harmonic traps has been measured through anisotropic
expansion \citep{Cao2012Science}, from which the locally uniform
shear viscosity has also been indirectly extracted \citep{Joseph2015PRL}.
The shear viscosity in Fig. \ref{fig16_USTCTransportCoefficients}(d)
turns out to be about two times larger than the previous data in the
superfluid phase. Therefore, the unitary Fermi gas appears to be more
viscous. As a quantitative measure, the viscosity-to-entropy ratio
$\eta/s$ is given in the inset, in units of the conjectured KSS low-bound
$\hbar/(4\pi k_{B})$ \citep{Kovtun2005PRL}. Around the transition,
$\eta/s$ is about 18 times larger than the bound, indicating that
a unitary Fermi superfluid is a good but not the perfect fluid as
we may anticipate.

\section{Second sound in interacting Bose gases}

First sound and second sound have also been measured in interacting
Bose gases, both in two dimensions near the Berezinskii--Kosterlitz--Thouless
(BKT) transition and in three dimensions, by Zoran Hadzibabic and
his colleagues at University of Cambridge \citep{Christodoulou2021Nature,Hilker2021arXiv}.
Here, we briefly introduce their observations.

\begin{figure}
\centering{}\includegraphics[width=1\columnwidth]{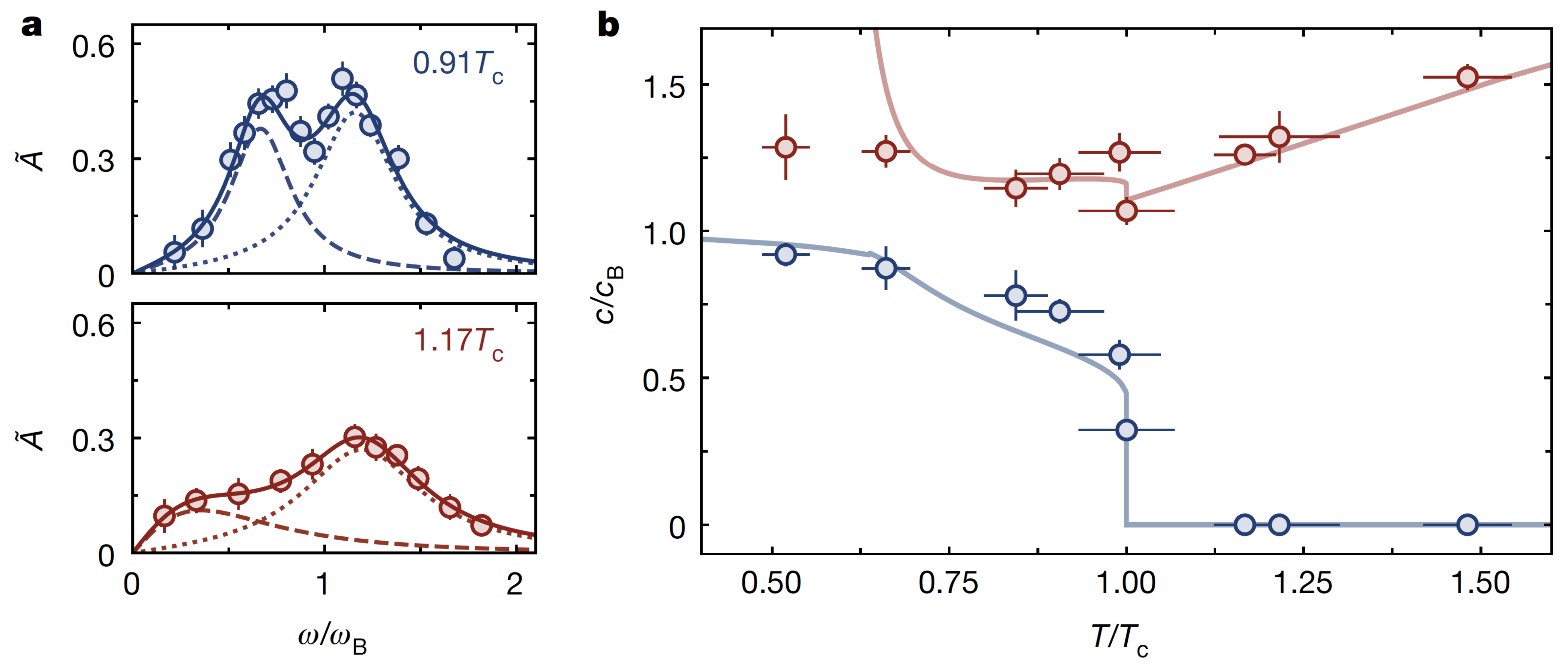}\caption{\label{fig17_BKT2D} (a) Normalized density response spectra $\tilde{A}(\omega)$
at $0.91T_{c}$ and $1.17T_{c}$. Below $T_{c}$ (top), two resonances
corresponding to the first (dotted) and second (dashed) sound are
observed. Above $T_{c}$ (bottom), the first-sound resonance (dotted)
exists, but the second sound is replaced by a diffusive, overdamped
mode (dashed). (b) Normalized sound velocities, $c_{1}/c_{B}$ (red)
and $c_{2}/c_{B}$ (blue), and the corresponding theoretical predictions
for infinite-systems without any free parameters. The discontinuities
in predicted velocities at $T_{c}$ are due to the jump in superfluid
density across the BKT transition. Adapted from Ref. \citep{Christodoulou2021Nature}.}
\end{figure}

\subsection{Second sound near the BKT transition}

In the 2D experiment \citep{Christodoulou2021Nature}, a 2D homogeneous
Bose gas of $^{39}$K atoms with surface density $n\simeq3\:\mu$m$^{-2}$
and area $L_{x}L_{y}$ is prepared in a node of a vertical 1D optical
lattice with a large harmonic trapping $\omega_{z}=2\pi\times5.5(1)$
kHz. Both the interaction and thermal energy per particle are below
$0.3\hbar\omega_{z}$, so the system is deep in the 2D regime. The
$s$-wave scattering length is increased by a Feshbach resonance to
$a=522(23)a_{0}$ with the Bohr radius $a_{0}$, so the dimensionless
2D interaction strength $\tilde{g}=\sqrt{8\pi}a/\sqrt{\hbar(m\omega_{z})}=0.64(3)$
is large enough to maintain the collisional hydrodynamic behavior.
There is a noticeable three-body loss due to the large interaction
strength, but the system appears to decay slowly without heating and
a steady-state may then be assumed.

At a given reduced temperature $T/T_{c}$, the Bose gas is perturbed
by an in-plane, spatially-uniformed force $F_{y}(t)=F_{0}\sin\left(\omega t\right)$
created by a magnetic field gradient $\Delta B$ along the $y$-direction.
The longest-wavelength sound mode(s) can then be excited, with wavenumber
$k=\pi/L_{y}$. The resulting density fluctuation would assume the
form, $\delta n(y,t)/n=b(t)\sin(ky)$, where the oscillating amplitude
$b(t)$ can be easily extracted from the displacement of the centre
of mass of the cloud $d(t)\equiv[R(\omega)\sin(\omega t)-A(\omega)\cos(\omega t)]\propto b(t)$.
The out-of-phase, absorptive response $A(\omega)$ then measures directly
the imaginary part of the density response function $\textrm{Im}\chi(k,\omega)$
at $k=\pi/L_{y}$.

In Fig. \ref{fig17_BKT2D}(a), the normalized absorptive response
$\tilde{A}(\omega)$ is reported both below and above the BKT transition.
The normalization is set according to the well-known $f$-sum rule
\citep{Griffin1993Book,Hu2010NJP}, $\int\omega\tilde{A}(\omega)d\omega=c_{B}^{2}k^{2}$,
with an appropriately chosen Bogoliubov sound velocity $c_{B}=\sqrt{\tilde{g}n}\hbar/m$
and Bogoliubov frequency $\omega_{B}=c_{B}k$ as the units. The data
have been fitted with two resonance functions, following the mode
decoupling in Eq. (\ref{eq:kappa1}) and Eq. (\ref{eq:kappa2}) (below
$T_{c}$) or Eq. (\ref{eq:kappa2p}) (above $T_{c}$). Below the BKT
transition at $T=0.91T_{c}$, two resonances with frequencies $\omega_{1}>\omega_{2}$
are clearly resolved, corresponding to the first sound and second
sound, respectively. The sound velocities extracted from the resonance
frequencies, i.e. $c_{1}=\omega_{1}/k$ and $c_{2}=\omega_{2}/k$,
are shown in Fig. \ref{fig17_BKT2D}(b), as a function of the reduced
temperature $T/T_{c}$. It is remarkable that the data can be well-explained
by the theoretical results based on Landau's two-fluid hydrodynamic
theory \citep{Ozawa2014PRL,Liu2014AoP,Ota2018PRA}, without any free
parameters. These predictions have been obtained using the accurate
equations of state and superfluid density from Monte Carlo simulations
\citep{Prokofev2002PRA}. Moreover, at the BKT transition, a discontinuity
in the second sound velocity can be clearly identified, due to the
sudden jump of superfluid density.

The sound attenuations of both first sound and second sound can also
be determined from the normalized absorptive response $\tilde{A}(\omega)$,
although they might be broadened by the loss-induced density drift.
From the resonance widths, it has been found $D_{1}=7(1)\hbar/m$
and $10(2)\hbar/m$ for the first sound below and above $T_{c}$,
respectively, and $D_{2}=6(1)\hbar/m$ for the second sound below
$T_{c}$. These sound diffusivities are several times larger than
that of strongly interacting quantum fluids (i.e., superfluid helium
or unitary Fermi gas).

\begin{figure}
\centering{}\includegraphics[width=1\columnwidth]{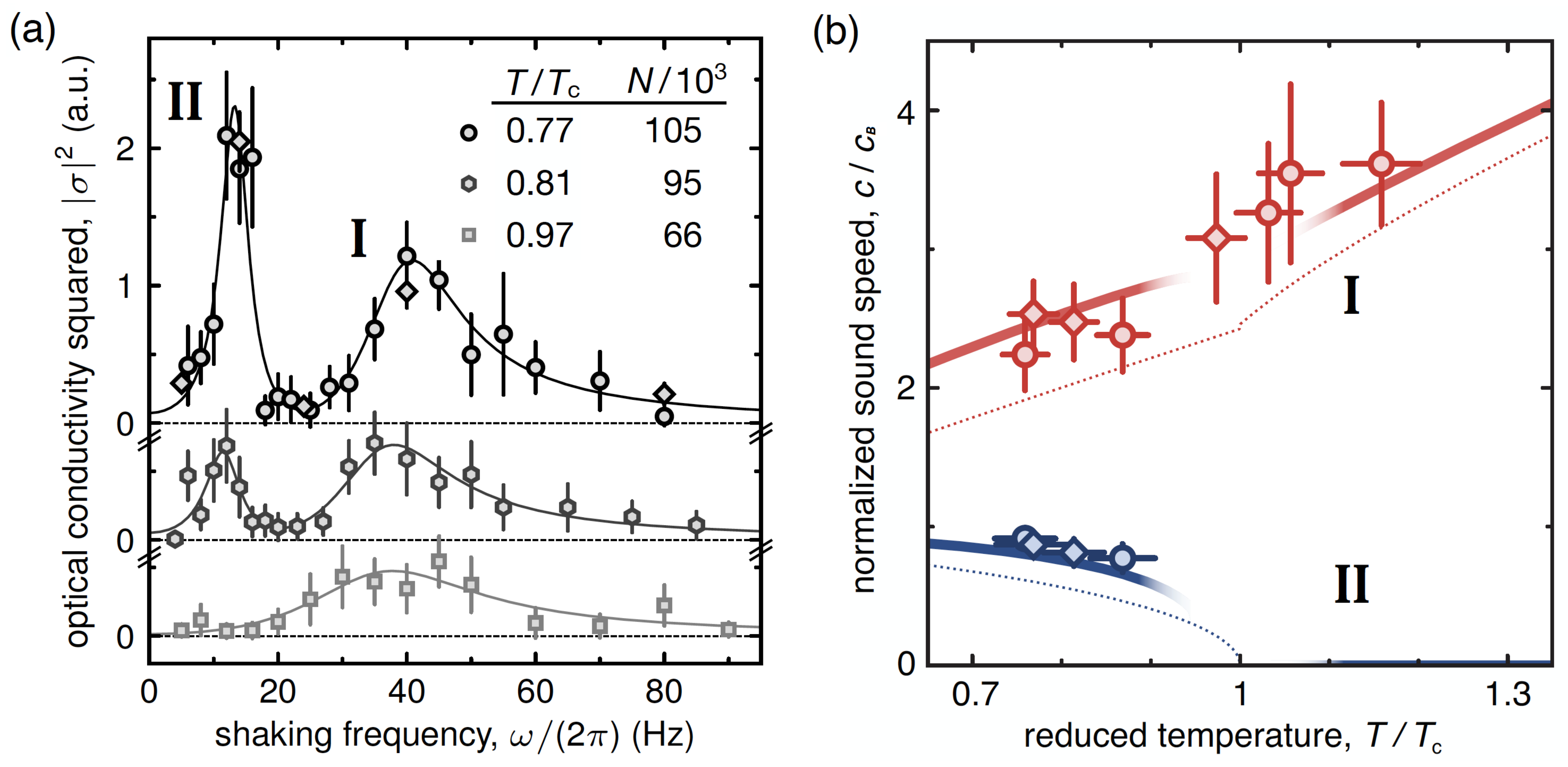}\caption{\label{fig18_BoseGas3D} (a) Optical conductivity squared $\left|\sigma\left(\omega\right)\right|^{2}$
at three different reduced temperatures. The two peaks correspond
to the two sound modes. The solid lines are fits using the sum of
two resonances. (b) First (red) and second (blue) sound velocities,
normalized by the Bogoliubov speed $c_{B}=\sqrt{4\pi na}\hbar/m$,
as a function of $T/T_{c}$. The solid lines are the fit-free predictions
of the two-fluid model, with $n_{s}$ calculated using the Popov mean-field
theory. The fade of the lines indicates the breakdown of the mean-field
theory near $T_{c}$. The dotted lines show the sound velocities calculated
by using non-interacting equation of state and condensate density
(as superfluid density). Adapted from Ref. \citep{Hilker2021arXiv}.}
\end{figure}

\subsection{Second sound in 3D compressible interacting Bose gases}

The above measurements have been recently extended to a 3D interacting
Bose gas confined in a cylindrical box trap of length $L=70(2)$ $\mu\textrm{m}$
in the $z$-direction and radius $R=9.2(5)$ $\mu\textrm{m}$ in the
transverse direction \citep{Hilker2021arXiv}. The total number of
atoms is $N=105(3)\times10^{3}$ and the $s$-wave scattering length
$a=480(20)a_{0}$, leading to a gas parameter $na^{3}\simeq0.93\times10^{-4}$.
Although the relatively large gas parameter enhances the three-body
loss and causes heating, the Bose cloud may still be treated as a
quasi-equilibrium system at the time scale of about 100 milliseconds.
The reach of the hydrodynamic regime is characterized by $kl_{\textrm{mfp}}\sim0.47<1$,
where the wavenumber $k=\pi/L$ for the lowest sound mode along the
axis and $l_{\textrm{mfp}}=(8\pi na^{2})^{-1}$ is the collisional
mean free path.

As in the 2D case, the lowest sound modes are excited by an axial
magnetic gradient that imposes a spatially uniform force along the
$z$-direction $F(t)=F_{0}\sin(\omega t)$. After a variable time
$t$, both the force and the trapping potential are turned off. The
axial 1D momentum distribution $n_{k}(t)$ is then measured after
a time-of-flight expansion of $30-45$ milliseconds. Consequently,
the center-of-mass velocity $v(t)=\hbar\left\langle k\right\rangle /m$
and current density $j(t)=nv(t)$ are determined. This provides a
complex optical conductivity $\sigma(\omega)\propto j/F\propto i\omega\chi(k,\omega)$
at $k=\pi/L$.

As shown in Fig. \ref{fig18_BoseGas3D}(a), well below $T_{c}$ (i.e.,
the top curve for $T=0.77T_{c}$), there are two well-resolved peaks
in the spectrum $\left|\sigma(\omega)\right|^{2}$, which correspond
to the first sound and second sound, respectively. The sound velocities
$c_{\textrm{I}}$ and $c_{\textrm{II}}$ can be directly read from
the peak positions $\omega_{\textrm{I}}$ and $\omega_{\textrm{II}}$,
i.e., $c_{\textrm{I,II}}=\omega_{\textrm{I,II}}/k$. As reported in
Fig. \ref{fig18_BoseGas3D}(b), the temperature dependence of the
two sound velocities can be well understood by applying a weak-coupling
mean-field theory (see also Fig. \ref{fig7_BECSoundVelocity}(a) for
a smaller gas parameter $na^{3}=10^{-5}$). The second sound attenuation
may also be extracted. However, the deduced amplitude-damping coefficient
does not show a clear dependence on the total number of atoms $N$.

\begin{figure}
\centering{}\includegraphics[width=1\columnwidth]{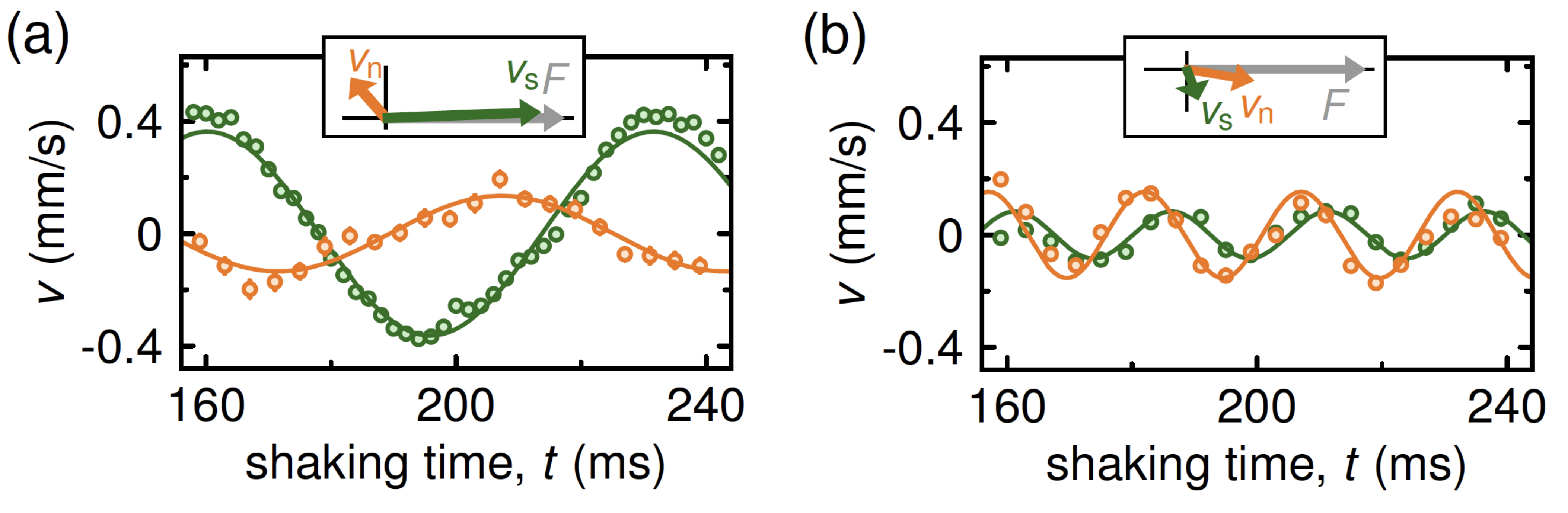}\caption{\label{fig19_BoseGas3D} Microscopic structure of the two sound modes
at $T/T_{c}=0.77$. (a) Extracted $v_{n}$ (orange) and $v_{s}$ (green)
for the second-sound resonance, exhibiting the out-of-phase motion
anticipated for second sound. (b) Analogous results for the first-sound
resonance. Adapted from Ref. \citep{Hilker2021arXiv}.}
\end{figure}

An interesting aspect of the 3D measurement is that the normal-fluid
velocity $v_{n}(t)$ can be extracted from the high-momentum wing
of the 1D momentum distribution $n_{k}(t)$. The superfluid velocity
$v_{s}(t)$ can then be deduced from the total current $j(t)$ based
on the theoretical superfluid fraction. This provides an unique way
to identify the microscopic structure of the second sound. As shown
in Fig. \ref{fig19_BoseGas3D}(a), at the second sound resonance the
extracted $v_{n}$ and $v_{s}$ are almost completely out-of-phase
as the time evolves, as predicted for second sound.

\section{Conclusions and outlooks}

In conclusions, the recent simultaneous observations of first sound
and second sound propagations in strongly interacting unitary Fermi
gases \citep{Sidorenkov2013Nature,Li2022Science} and in interacting
Bose gases \citep{Christodoulou2021Nature,Hilker2021arXiv} provide
the textbook cases of the celebrated Landau's two-fluid hydrodynamic
theory \citep{Landau1941PR,Khalatnikov2000Book}. Unlike superfluid
helium, both interacting Fermi gases and Bose gases are highly tunable
and controllable systems \citep{Bloch2008RMP}, with the corresponding
microscopic model Hamiltonians well established. Therefore we now
have an excellent toolbox to better understand the microscopic details
of superfluidity and the superfluid phase transition in the strongly
interacting regime.

In this respect, improved measurements of the unitary Fermi gas near
the superfluid phase transition are of particular interest. In liquid
helium, the investigations of critical divergences in the thermal
conductivity $\kappa$ above the $\lambda$-transition and in the
second-sound diffusivity $D_{2}$ below the $\lambda$-transition
play a vital role in setting up the effective dynamic scaling theory
for the critical mode across the superfluid transition \citep{Ferrell1967PRL,Hohenberg1977RMP}.
For a homogeneous unitary Fermi gas, the recent first measurement
of both $D_{2}$ and $\kappa$ at USTC already indicates a much larger
quantum critical regime, which is about 100 times larger than that
of liquid helium. A systematic exploration of this large quantum critical
region with improved temperature controllability could be achieved
in the near future, paving the way to map out several long-sought
universal critical dynamic scaling functions \citep{Hohenberg1977RMP}.

Future experiments on interacting Bose gases are equally important,
as they present highly compressible examples of quantum liquids. However,
the use of a large $s$-wave scattering length will cause three-body
loss and heating, which prevent the precise determination of the sound
velocities and sound attenuations. To overcome this difficulty, we
may consider a molecular Bose-Einstein condensate, created by a two-component
interacting Fermi gas on the BEC side. In particular, by adjusting
the box-trap geometry and optimizing the system, the USTC setup in
Fig. \ref{fig13_USTCsetup}(a) can be readily modified to realize
a two-dimensional homogeneous Fermi superfluid. This provides an ideal
platform for investigating the first and second sound attenuation
and related quantum transport across the BKT transition \citep{Tononi2021PRA,Bighin2016PRB,Mulkerin2017PRA},
without three-body loss.

\begin{figure}
\centering{}\includegraphics[width=1\columnwidth]{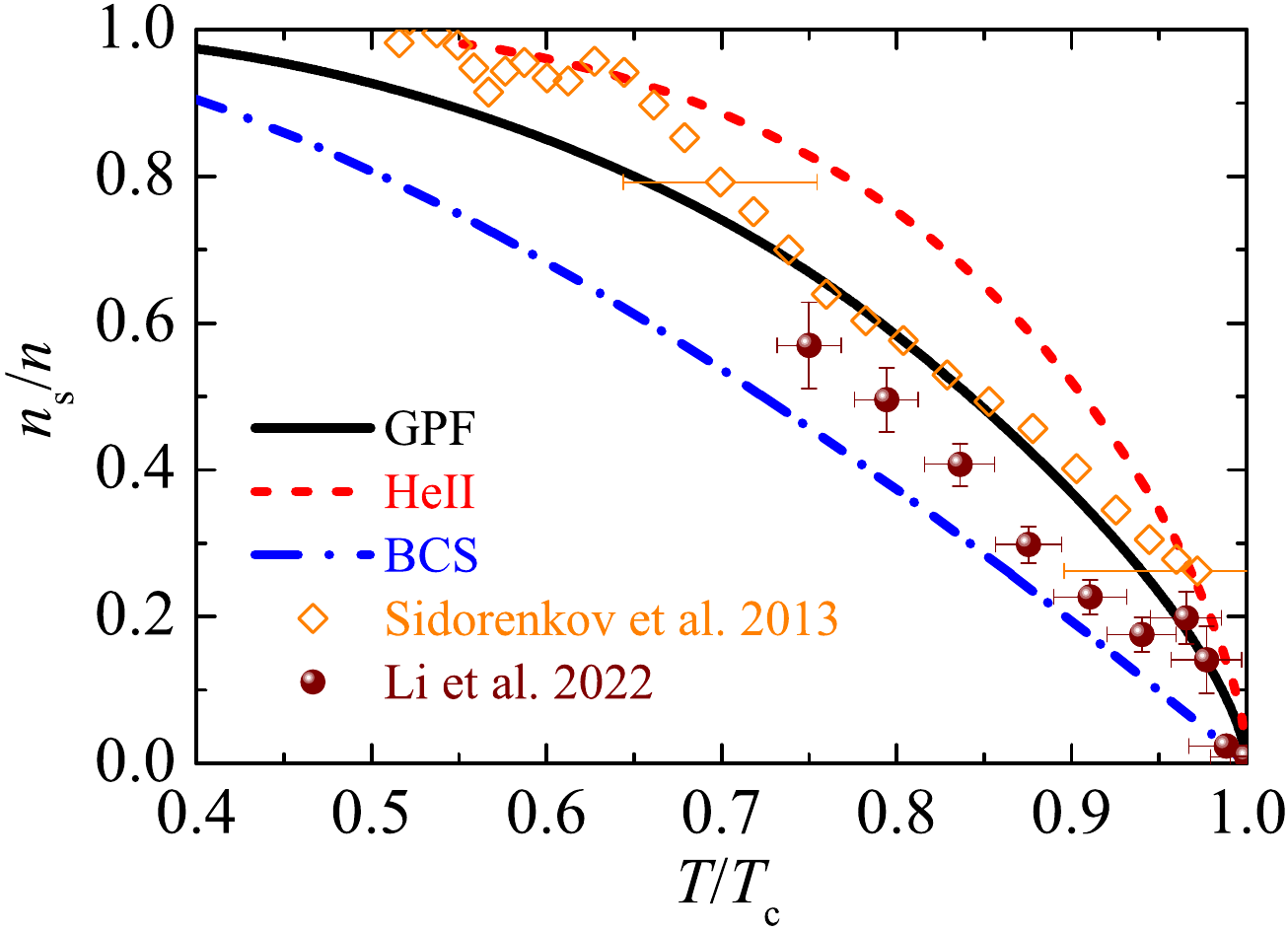}\caption{\label{fig20_nsUnitaryFermiGas} The superfluid fraction of a unitary
Fermi gas measured from the Insbruck experiment in 2013 \citep{Sidorenkov2013Nature}
(empty diamonds) and from the USTC experiment in 2022 (solid circles)
\citep{Li2022Science}, as a function of the reduced temperature $T/T_{c}$.
The data are compared with the theoretical predictions from the Gaussian
pair fluctuation (GPF) theory (black solid line) \citep{Fukushima2007PRA,Taylor2008PRA}
and mean-field BCS theory (blue dot-dashed line). The superfluid fraction
of superfluid helium is also shown by a red dashed line.}
\end{figure}

\begin{figure}
\centering{}\includegraphics[width=1\columnwidth]{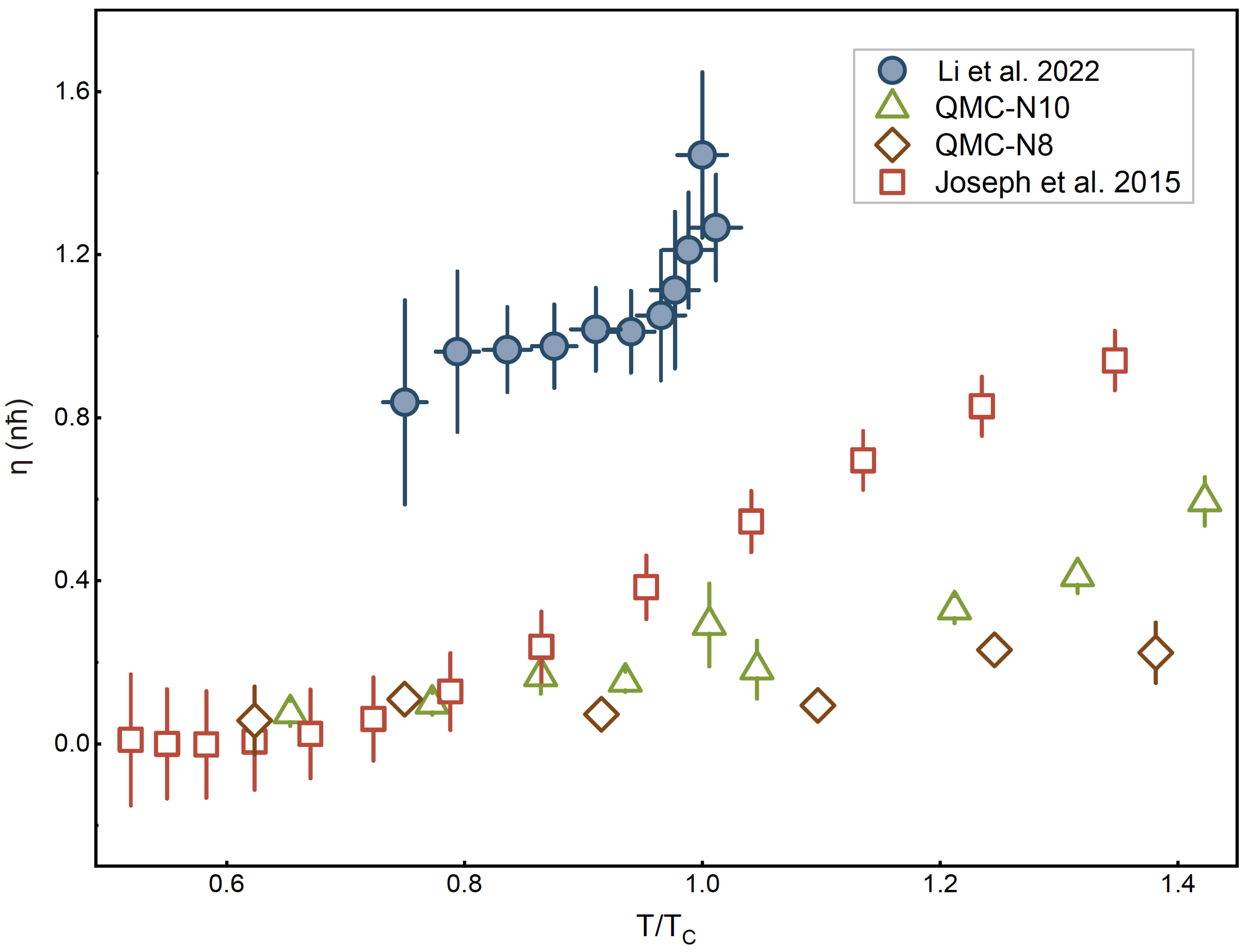}\caption{\label{fig21_ShearViscosityUFG} The shear viscosity of a unitary
Fermi gas measured at NCSU in 2015 (empty squares) \citep{Joseph2015PRL}
and at USTC group in 2022 (solid circles) \citep{Li2022Science},
as a function of the reduced temperature $T/T_{c}$. For comparison,
we show also the quantum Monte Carlo simulations with lattice size
$N=8^{3}$ (empty diamonds) and $N=10^{3}$ (empty triangles) \citep{Wlazlowski2013PRA}.
From Ref. \citep{Li2022Science}.}
\end{figure}

\subsection{Many-body challenges}

The milestone experiments reviewed in this work already present a
number of intriguing and challenging many-body problems for theorists.
For example, both the superfluid density and the transport coefficients
(shear viscosity $\eta$ and thermal conductivity $\kappa$) of the
unitary Fermi gas are notoriously to calculate theoretically. As shown
in Fig. \ref{fig20_nsUnitaryFermiGas}, the new data on the superfluid
fraction from the USTC experiment lie between the theoretical predictions
from the Gaussian pair fluctuation theory and the standard BCS mean-field
theory. New calculations are definitively needed, to better explain
the experimental observation (for a recent attempt, see Ref. \citep{UnitarySuperfluidDensity2022arXiv}).
In Fig. \ref{fig21_ShearViscosityUFG}, the USTC results for the shear
viscosity (blue circles) are compared with the local shear viscosity
extracted from the trap-averaged shear viscosity \citep{Joseph2015PRL}
and the QMC calculations \citep{Wlazlowski2013PRA}. These results
seem not to reach an agreement. Improved theoretical calculations
of the shear viscosity of a unitary Fermi superfluid, based on either
strong-coupling diagrammatic theories or QMC simulations, are timely
required to resolve the discrepancy.

\begin{figure}[t]
\centering{}\includegraphics[width=1\columnwidth]{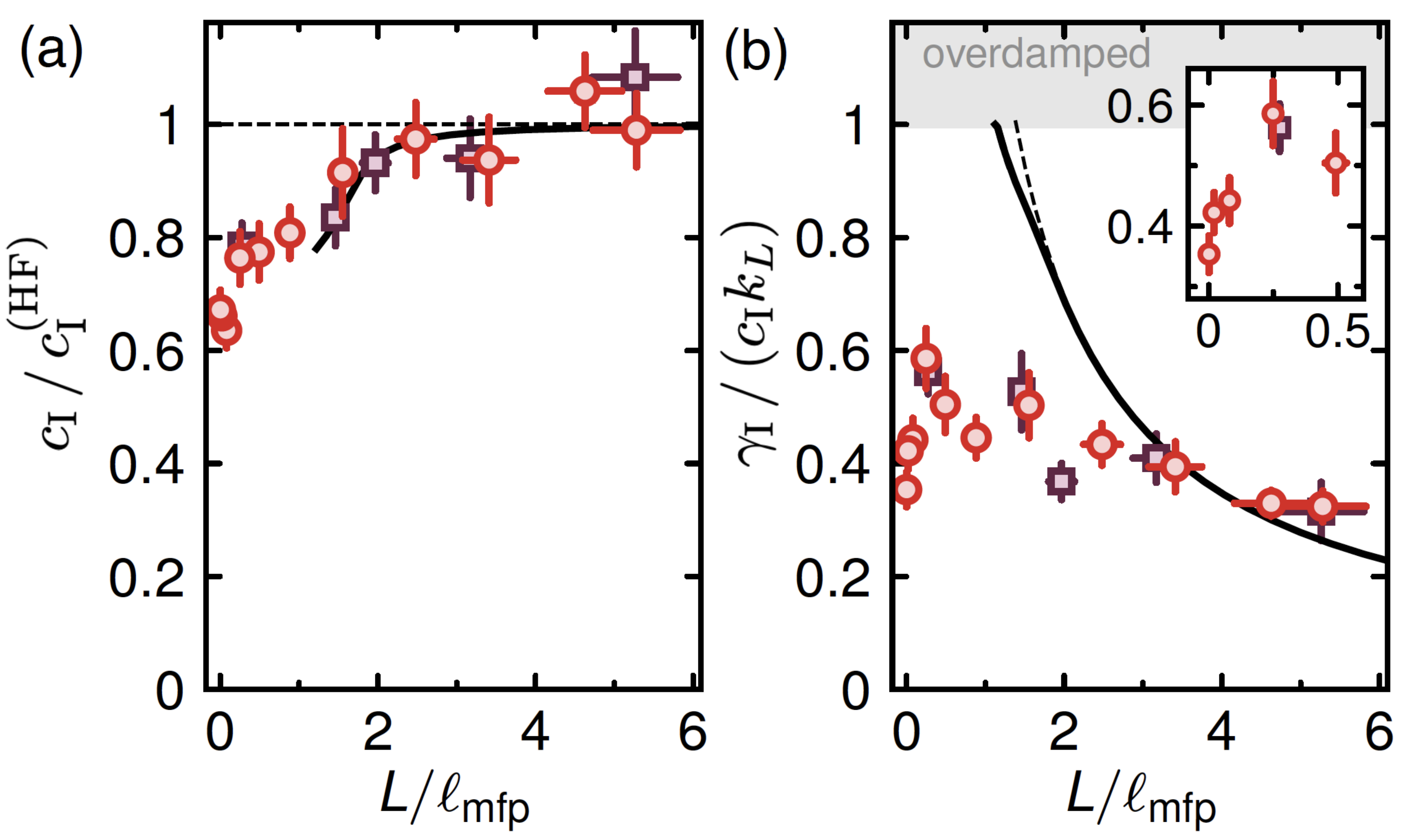}\caption{\label{fig22_CollisionlessToHydrodynamicBoseGas3D} The first sound
velocity (a) and its relative damping rate (b) of a 3D interacting
Bose gas as a function of the inverse Knudsen number $L/l_{\textrm{mfp}}$
in its normal state at $T\simeq1.3T_{c}$. The first sound velocity
is measured in units of $c_{\textrm{HF}}$, given in Eq. (\ref{eq:cHF}).
For large $L/l_{\textrm{mfp}}$, the data approach the results of
the full hydrodynamic calculation (solid lines). The inset in (b)
highlights the damping in the collisionless limit. Adapted from Ref.
\citep{Hilker2021arXiv}.}
\end{figure}

\subsection{From the collisionless to hydrodynamic regime}

On the other hand, less is known about the transition from the hydrodynamic
regime to the collisionless regime. For unitary Fermi gas, the USTC
setup can be easily extended to investigate the temperature and wave
number dependence of the density response. As a result, the transition
from collisionless to hydrodynamic behavior could be fully characterized
and thus illuminate the establishment of hydrodynamics in the strongly
interacting regime. We note that, in the normal state the first sound
density response at different wavenumbers has been recently investigated
by the NCSU group \citep{Baird2019PRL,Wang2022PRL}. The sound velocity
and damping rate of zero sound in the collisionless regime have been
studied by several experimental groups \citep{Kuhn2020PRL,Biss2022PRL}
and have been theoretically addressed \citep{Castin2017PRL,Zou2018PRA}.

For a compressible weakly interacting Bose gas, the collisionless
to hydrodynamic transition in the normal state has been experimentally
explored \citep{Hilker2021arXiv}. As reported in Fig. \ref{fig22_CollisionlessToHydrodynamicBoseGas3D},
the transition is tuned by the inverse Knudsen number $L/l_{\textrm{mfp}}$
for the lowest excitation mode with $k=\pi/L$. It is readily seen
that, by increasing $L/l_{\textrm{mfp}}$ above 3 or decreasing $kl_{\textrm{mfp}}$
down to 1, both the (first) sound velocity and its damping rate approach
the predictions given by the full hydrodynamic calculations (solid
lines). For such a three-dimensional weakly interacting Bose gas,
it would be interesting to develop a quantitative microscopic description
of the transition from the hydrodynamic regime to the collisionless
regime \citep{Griffin2009Book}. In two dimensions close to the BKT
transition, the sound propagation in an interacting Bose gas has also
been experimentally observed by J.\LyXThinSpace L. Ville et al. \citep{Ville2018PRL}
and theoretically analyzed by using the random phase approximation
\citep{Ota2018PRL} and the collisionless kinetic theory \citep{Cappellaro2018PRA}.
\begin{acknowledgments}
This research was supported by the Australian Research Council's (ARC)
Discovery Program, Grant No. DP180102018 (X.-J.L) and by the National
Natural Science Foundation of China (NSF-China), Grant No. 11874340
(X.-C. Y). Xia-Ji Liu was also supported in part by the National Science
Foundation under Grant No. PHY-1748958, during her participation in
the KITP program ``Living Near Unitarity'' and the KITP conference
``Opportunities and Challenges in Few-Body Physics: Unitarity and
Beyond''.
\end{acknowledgments}

\bibliography{Ref2ndSoundReview}

\end{document}